\documentclass[10pt,a4paper,twocolumn]{IEEEtran}
\usepackage[comma,numbers,square,sort&compress]{natbib}
\usepackage{algorithm} 
\usepackage{algorithmic} 
\usepackage{multirow} 
\usepackage{xcolor}
\usepackage{amsmath}
\usepackage{amsmath}
\usepackage{multirow}
\usepackage{latexsym}
\usepackage{setspace}
\usepackage{url}
\usepackage{subfigure}
\usepackage{fancyhdr}
\usepackage{amsmath}
\usepackage{multirow}
\usepackage{amssymb }
\usepackage{color}
\usepackage{graphics} 
\usepackage{graphicx}
\usepackage{pmat}
\usepackage{arydshln}

\newtheorem{theorem}{\textbf{Theorem}}
\newtheorem{lemma}{\textbf{Lemma}}
\newtheorem{corollary}{\textbf{Corollary}}
\newtheorem{remark}{\textbf{Remark}}
\newtheorem{definition}{\textbf{Definition}}

\begin{document}

\title{\LARGE \bf Affine Dependence of Network Observability/Controllability on Its Subsystem Parameters and Connections}


\author{Tong Zhou$^{\dag}$ and Yuyu Zhou 
\thanks{This work was supported in part by the NNSFC under Grant 61733008 and 61573209. This work has been submitted to the IEEE for possible publication. Copyright may be transferred without notice, after which this version may no longer be accessible.}
\thanks{Tong Zhou$^{\dag}$ (corresponding author) and Yuyu Zhou are with the Department of Automation, Tsinghua University, Beijing, 100084, P.~R.~China
        {(email: {\tt\small tzhou@mail.tsinghua.edu.cn, zhouyy18@mails.tsinghua.edu.cn}).}}%
}
\maketitle

\begin{abstract} This paper investigates observability/controllability of a networked dynamic system (NDS) in which system matrices of its subsystems are expressed through linear fractional transformations (LFT). Some relations have been obtained between this NDS and descriptor systems about their observability/controllability. A necessary and sufficient condition is established with the associated matrices depending affinely on subsystem parameters/connections. An attractive property of this condition is that all the required calculations are performed independently on each individual subsystem. Except well-posedness, not any other conditions are asked for subsystem parameters/connections. This is in sharp contrast to recent results on structural observability/controllability which is proven to be NP hard. Some characteristics are established for a subsystem which are helpful in constructing an observable/controllable NDS. It has been made clear that subsystems with an input matrix of full column rank are helpful in constructing an observable NDS, while subsystems with an output matrix of full row rank are helpful in constructing a controllable NDS. These results are extended to an NDS with descriptor form subsystems. As a byproduct, the full normal rank condition of previous works on network observability/controllability has been completely removed. On the other hand, satisfaction of this condition is shown to be appreciative in building an observable/controllability NDS.
\end{abstract}


\section{Introduction}

In system designs, it is essential to at first build a plant capable of reaching good performances. When a networked dynamic system (NDS) is to be designed, this problem is related to both selecting subsystem parameters and designing subsystem connections \cite{ixkm2010,boq2017,sbkkmpr11,zyl18}. To achieve this objective, some explicit relations seem necessary between system achievable performances and its parameters/connections. On the other hand, observability/controllability is essential for a system to properly work, noting that they are closely related to various important system properties. Examples include fault detection, optimal control, pole placement, state estimation, etc. It is widely believed that an uncontrollable/unobservable plant can hardly be anticipated to have satisfactory regulation/estimation performances  \cite{ksh00,sbkkmpr11,zdg96,zyl18}.

Observability and controllability are now well developed concepts in system analysis and synthesis, and various criteria have been established, such as the PBH test, controllability/observability matrix, etc. \cite{ksh00,zdg96,zyl18}. In addition, these concepts have been studied from various distinctive aspects. For example, it is now extensively known that for various types of systems, controllability and observability are generic system properties. That is, rather than numerical values of system matrices, it is the state connections, as well as the connections from an input to the system states (the connections from the system states to an output), that determine the controllability (observability) of a system. Motivated by these observations, structural controllability/observability is developed and studied by many researchers \cite{dcv03,Lin74,zyl18}. \cite{ccvfb12} and \cite{zhou18} reveal that the minimal input/output number guaranteeing the structural controllability/observability of an NDS is determined by its subsystem dynamics. The problem is proved to be NP-hard in \cite{Olsehvsky14} of searching the sparest output/input matrix with the associated NDS observable/controllable. \cite{pzb14} suggests some metrics for  quantitatively analyzing hardness in controlling a system. It has been made clear in \cite{zhou18} that to construct an observable/controllable NDS, each subsystem should be observable/controllable. Another observation there is that the minimal number for subsystem outputs/inputs is equal to the maximal geometric multiplicity of its state transition matrix. And so on.

While various results have been obtained for NDS observability/controllability, many important issues remain unsolved. Influences of subsystem connections/dynamics etc., on the observability/controllability of the whole system, are some examples \cite{cam14,cpa17,zyl18}. In addition, technology developments, stringent performance requirements, etc. are rapidly producing complicated systems with an increasing number of subsystems that have distinguished working mechanics \cite{ixkm2010,boq2017,zhou15}. For these systems, numerical stability and computational costs are essential issues with great challenges \cite{Olsehvsky14,sbkkmpr11,zyl18}.

When system matrices of each subsystem can be expressed as an linear fractional transformation (LFT) of its first principle parameters (FPP), NDS structural controllability has been recently studied in \cite{zz19}. It has been shown that controllability for these systems is a generic property and their structural controllability verification is in general NP hard. This is quite discouraging, noting that LFT is a very effective expression in system analysis and synthesis, and various systems have  the property that although elements of its system matrices are not algebraically independent of each other, they can be written as an LFT of its algebraically independent FPPs \cite{zdg96,zyl18}.

To settle the issue of NDS observability/controllability verifications under the situation that system matrices of its subsystems are expressed by LFTs, in this paper, rather than their structural counterparts, observability and controllability are directly studied. Surprisingly, it has been observed that this verification problem can be converted to the verification of the controllability/observability of a particular descriptor system. By means of the Kronecker canonical form (KCF) of a matrix pencil, a rank based condition is derived with the associated matrix depending affinely on both subsystem parameters and connections. This condition keeps the attractive properties of the verification procedure reported in \cite{zhou15,zyl18} that in obtaining the associated matrices, the involved numerical calculations are performed independently on each subsystem, which make its verification scalable for a large scale NDS. In deriving these results, except a well-posedness condition, not any other requirements are asked for a subsystem FPP or a subsystem connection. This explicit relation with  subsystem parameters/connections seem useful in designing system topologies and selecting subsystem parameters. A byproduct of this investigation is that the full normal rank condition asked in \cite{zhou15,zyl18} has been completely removed. On the other hand, it has been made clear that satisfaction of this rank condition by a subsystem is appreciative in reducing difficulties of constructing a controllable/observable NDS.

On the basis of this condition, it is shown that a subsystem with an input matrix of full column rank (FCR) is helpful in constructing an observable NDS that receives signals from other subsystem, while a subsystem with an output matrix of full row rank (FRR) is helpful in constructing a controllable NDS that sends signals to other subsystems. Some rank conditions have also been derived for a subsystem with which an observable/controllable NDS can be constructed more easily, and these conditions can be independently verified for each subsystem. These results are expected to be helpful in studying optimal sensor/actuator placements. Extensions to an NDS with descriptor form subsystems have also been dealt with. Similar results have been obtained.

The above results are in sharp contrast to those about structural controllability/observability, and make controllability/observability verification in principle feasible for a large scale NDS with LFT parameterized system matrices for each subsystem, noting that in \cite{zz19}, structural controllability verification has been proven to be NP hard for these systems, and an effective verification algorithm is developed only for the case in which the matrix has a diagonal parametrization that is constructed from all subsystem FPPs and the SCM, which appears to be a great restriction on the applicability of the obtained results to a practical problem.

The remaining of this paper is structured as follows. The next section gives an NDS model and some preliminary results.
NDS observability/controllability is dealt with in Section III. An application of these results to sensor/actuator placements are discussed in Section IV, while Section V investigates how to extend them to an NDS when the dynamics of some subsystems are described by a descriptor form. A numerical example is provided in Section VI to illustrate the obtained theoretical results. Finally, conclusions are given in Section VII  which reveals some further issues. Four appendices are attached to provide proofs of several technical results.

The following symbols and notation are adopted. ${\cal R}^{n}$ and $\cal C$ represent respectively the $n$ dimensional real Euclidean space and the set of complex numbers. ${\rm\bf det} \left(\cdot\right)$  stands for the determinant of a square matrix, while $\cdot^{\perp}$ the matrix whose columns form a base of the null space of a matrix. ${\rm\bf
diag}\!\{X_{i}|_{i=1}^{L}\}$ denotes a block diagonal matrix with
its $i$-th diagonal block being $X_{i}$, while ${\rm\bf
col}\!\{X_{i}|_{i=1}^{L}\}$ the matrix/vector stacked by
$X_{i}|_{i=1}^{L}$ with its $i$-th row block matrix/vector being
$X_{i}$. When $A$ is an $m\times n$ dimensional matrix and $k$/$\cal K$ an element/subset of $\{1,2,\cdots,n\}$, $A(1:k)$ represents  the matrix constructed from its first $k$ columns, while $A({\cal K})$ the matrix constructed from its columns indexed by the set $\cal K$. $0_{m\times n}$ and $0_{m}$ stand respectively for the $m\times n$ dimensional zero matrix and the $m$ dimensional zero column vector. Their subscripts are often  omitted when the omission does not cause confusions. Superscripts $T$ and $H$ are adopted to represent the transpose and the conjugate transpose
of a vector/matrix respectively. If at some values of its variable(s), the value of a matrix valued function is of FRR/FCR, it is said to be of full normal row/column rank (FNRR/FNCR).

Results of Section III in this paper have been presented without proof in the 58th IEEE Conference on Decision and Control \cite{zhou19}. In addition to providing proofs of these results, this paper also investigates NDS actuator/sensor placements, as well as NDS controllability/observability verifications when subsystem dynamics are described through a descriptor form.

\section{NDS Model and Preliminary Results}

In real world problems, NDS subsystems usually do not have identical input-output relations. In \cite{zhou15,zyl18}, an approach is suggested to describe the dynamics of a general linear time invariant (LTI) NDS. In this paper, to clarify relations between subsystem FPPs and its system matrices, the following model is adopted to describe the dynamics of the $i$-th subsystem ${\bf{\Sigma}}_i$ of an NDS $\bf \Sigma$ constituted from  $N$ subsystems. This model is also utilized in \cite{zz19} for continuous time NDSs.
\begin{equation}
\hspace*{-0.25cm}\begin{array}{l}
\left[\!\! {\begin{array}{*{20}{c}}
{\delta({{x}}(t,i))}\\
{{z}(t,i)}\\
{{y}(t,i)}
\end{array}}\!\! \right] \!=\! \left\{\! {\left[\!\! {\begin{array}{*{20}{c}}
{A_{\rm\bf xx}^{[0]}(i)} & {A_{\rm\bf xv}^{[0]}(i)} & {B_{\rm\bf x}^{[0]}(i)}\\
{A_{\rm\bf zx}^{[0]}(i)} & {A_{\rm\bf zv}^{[0]}(i)} & {B_{\rm\bf z}^{[0]}(i)}\\
{C_{\rm\bf x}^{[0]}(i)} & {C_{\rm\bf v}^{[0]}(i)} & {D^{[0]}(i)}
\end{array}}\!\! \right] \!\!+\!\! \left[ \!\!{\begin{array}{*{20}{l}}
{H_1{(i)}}\\
{H_2{(i)}}\\
{H_3{(i)}}
\end{array}} \!\!\right]\!\!\times } \right.\\
\left. {\left. {\begin{array}{*{20}{l}}
 \\
 \\
\end{array}} \right.{P(i)}{{\left[I - {G{(i)}}{P{(i)}}\right]}^{ - 1}}\left[\!\! {\begin{array}{*{20}{c}}
{F_1{(i)}} & {F_2{(i)}} & {F_3{(i)}}
\end{array}}\!\! \right]}\! \right\}\!\!\left[\!\! {\begin{array}{*{20}{c}}
{{x}(t,i)}\\
{{v}(t,i)}\\
{{u}(t,i)}
\end{array}}\!\! \right]
\end{array}
\label{eqn:1}
\end{equation}
Here, $\delta(\cdot)$ denotes either the derivative of a function with respect to time or a forward time shift operation. In other words, the above model can be either continuous time or discrete time. Moreover, $t$ stands for the temporal variable, $x(t,i)$ the state vector of this subsystem, $y(t,i)$ and $u(t,i)$ its external output and input vectors respectively,  $z(t,i)$ and $v(t,i)$ its internal output and input vectors respectively, representing signals sent to other subsystems and signals gotten from other subsystems. All subsystem parameters are given in the matrix $P(i)$, which may be a concentration or a reaction ratio in biological/chemical processes, a resistor, an inductor or a capacitor in electrical/electronic systems, a mass, a spring or a damper in mechanical systems, etc. They are usually called FPPs (first principle parameter), and can be chosen/tuned in system designs. The matrices $G{(i)}$, $H_j{(i)}|_{j=1}^{3}$ and $F_j{(i)}|_{j=1}^{3}$ are introduced to indicate how the system matrices of this subsystem is changed by its FPPs. These matrices, as well as the matrices
${A_{\rm\bf *\#}^{[0]}(i)}$, ${B_{\rm\bf *}^{[0]}(i)}$, ${C_{\rm\bf *}^{[0]}(i)}$ and ${D^{[0]}(i)}$, in which ${\rm\bf *,\#}={\rm\bf x}$, ${\rm\bf u}$, ${\rm\bf v}$, ${\rm\bf y}$ or ${\rm\bf z}$, are often used to
represent chemical, biological, physical or electrical principles governing subsystem dynamics, such as Netwon's mechanics, the Kirchhoff's current law, etc., which implies that they are usually prescribed and can hardly be chosen or tuned in designing a system.

In this model, the matrix $P(i)$ with $i\in\{1,2,\cdots,N\}$ is in principle constituted from fixed zero elements and FPPs of Subsystem ${\bf{\Sigma}}_i$. In some applications, a simple FPP function may be more convenient, such as the square root of a FPP, the multiplication of some  FPPs, etc. These transformations do not change conclusions in this paper, provided that they form a bijective global transformation. To avoid awkward statements, they are called pseudo FPPs (PFPP) in this paper, and are often  assumed to be independent of each other algebraically.

Compared with the subsystem model utilized in \cite{zhou15,zyl18},
it is believed that the above model is closer to actual input-output relations of a dynamic plant, due to that each system matrix, that is, ${A_{\rm\bf *\#}(i)}$, ${B_{\rm\bf *}(i)}$, ${C_{\rm\bf *}(i)}$ and ${D(i)}$, in which ${\rm\bf *,\#}={\rm\bf x}$, ${\rm\bf u}$, ${\rm\bf v}$, ${\rm\bf y}$ or ${\rm\bf z}$, is represented as a matrix valued function of the parameter matrix $P(i)$, which reflects the well known fact that in an actual system, elements in its system matrices are usually not algebraically independent of each other, and some of them can even not be tuned in system designs. Noting that an LFT is capable of representing any rational functions \cite{zdg96}, it is believed that a large class of systems can be described by the aforementioned model. A more detailed discussion can be found in \cite{zz19} on engineering motivations of the aforementioned model. To have a concise presentation, the dependence of a system matrix of the subsystem ${\bf{\Sigma}}_i$ on its parameter matrix $P(i)$ is usually not explicitly expressed, except when this omission may cause some significant confusions.

When effects from part of the subsystem FPPs on the performances of the whole NDS are to be investigated, the aforementioned model can also be applied. This can be done simply through prescribing the other FPPs to some  specific numerical values.

Denote vectors ${{\rm{{\bf{{\rm {col}}}}}}\{z(t,i)|_{i=1}^N\}}$ and ${{\rm{{\bf{{\rm {col}}}}}}}\{v(t,i)|_{i=1}^N\}$ respectively by $z(t)$ and $v(t)$. In this paper, it is assumed as in \cite{zhou15,zyl18} that NDS subsystem interactions are described by
\begin{equation}
v(t)=\Phi z(t)
\label{eqn:2}
\end{equation}
The matrix $\Phi$ is called subsystem connection matrix (SCM), which describes influences among different NDS subsystems. If each subsystem  is regarded as a node and each nonzero element of its SCM as an edge, a graph can be constructed for an NDS, which is usually called the structure or topology of the corresponding NDS.

Throughout this paper, the following assumptions are adopted.

\begin{itemize}
\item The vectors $u(t,i)$, $v(t,i)$, $x(t,i)$, $y(t,i)$ and $z(t,i)$  respectively have a dimension of $m_{{\rm\bf u}i}$, $m_{{\rm\bf v}i}$, $m_{{\rm\bf x}i}$, $m_{{\rm\bf y}i}$ and $m_{{\rm\bf z}i}$.
\item Every NDS subsystem ${\bf{\Sigma}}_i$ with $i\in\{1,2,\cdots,N\}$ is well-posed.
\item The NDS ${\bf{\Sigma}}$ itself is well-posed.
\end{itemize}

The first assumption is introduced to clarify vector size, while well-posedness of a system means that its states respond solely to each pair of their initial values and external inputs, that is necessary for a system to properly work  \cite{ksh00,sbkkmpr11,zdg96,zyl18}. This means that all these three assumptions should be met by a practical system. In other words, the assumptions adopted here are not quite restrictive.

By means of the above symbols, integers $M_{{\rm\bf x}i}$, $M_{{\rm\bf
v}i}$, $M_{\rm\bf x}$ and $M_{\rm\bf v}$ are defined respectively as $M_{\rm\bf
x}={\sum_{k=1}^{N} m_{{\rm\bf x}k}}$, $ M_{\rm\bf v}={\sum_{k=1}^{N}
m_{{\rm\bf v}k}}$, and $M_{{\rm\bf x}i}=M_{{\rm\bf v}i}=0$ when
$i=1$, $M_{{\rm\bf x}i}={\sum_{k=1}^{i-1} m_{{\rm\bf x}k}}$,
$M_{{\rm\bf v}i}={\sum_{k=1}^{i-1} m_{{\rm\bf v}k}}$ when $2\leq
i\leq N$. Then the dimension of the SCM $\Phi$ is $M_{\rm\bf v}\times M_{\rm\bf z}$. When this matrix is partitioned consistently with the dimensions of the vectors $z(t,i)|_{i=1}^{N}$ and $v(t,i)|_{i=1}^{N}$, its $i$-th row $j$-th column block, denoted by $\Phi_{ij}$, has a dimension of $m_{{\rm\bf v}i}\times m_{{\rm\bf z}j}$.  This submatrix reflects direct effects of Subsystem ${\bf{\Sigma}}_j$ on Subsystem ${\bf{\Sigma}}_i$, $i,j=1,2,\cdots,N$.

Briefly, in system analysis and synthesis, a system is called controllable if there exists an external input vector that can maneuver its sate vector from any prescribed initial value to any prescribed final value, and it is called observable if the value of its initial state vector can be recovered from the time history of its external input/output vectors. It is well known that controllability and observability are dual properties of a system, which means that observability of a system is equivalent to the controllability of its transpose, and vice versa \cite{ksh00,zdg96}. These characteristics keep valid for an NDS \cite{zhou15,zyl18}. 

To develop a computationally feasible condition for NDS observability/controllability verifications, the following well known results of matrix analyses are introduced \cite{Gantmacher59,hj91}.

\begin{lemma}
Divide a matrix $A$ as $A = \left[ A_{1}^{T} \;  A_{2}^{T}\right]^{T}$, and assume that $A_{1}$ is not of FCR. Then the matrix $A$ is of FCR, if and only if the matrix $A_{2}A_{1}^{\perp}$ is.
\label{lemma:3}
\end{lemma}

When the matrix $A_{1}$ is of FCR, $A_{1}^{\perp}=0$. In this case, for an arbitrary matrix $A_{2}$ with a compatible dimension, the matrix $A = \left[ A_{1}^{T} \;  A_{2}^{T}\right]^{T}$ is of FCR obviously.

Based on these conclusions, the following results are derived. They are greatly helpful in exploiting block diagonal structures of the matrices in NDS observability/controllability verifications.

\begin{lemma}
Assume that $A_{i}|_{i=1}^{3}$ and $B_{i}|_{i=1}^{3}$ are some matrices having compatible dimensions, and the matrix $A_{2}$ is of FCR. Then the matrix
$\left[\begin{array}{c}
{\rm\bf diag}\left\{A_{1},\;A_{2},\; A_{3}\right\} \\ \left[ B_{1} \;\;\;\; B_{2} \;\;\;\; B_{3}\right]\end{array}\right]$ is of FCR, if and only if the
matrix
$\left[\begin{array}{c}
{\rm\bf diag}\left\{A_{1},\; A_{3}\right\} \\ \left[ B_{1} \;\; \;\; B_{3}\right]\end{array}\right]$ is.
\label{lemma:6}
\end{lemma}

\hspace*{-0.4cm}{\it\bf Proof:} Note that
\begin{eqnarray}
\left[\!\begin{array}{ccc}
A_{1} & 0 & 0 \\ 0 & A_{2} & 0 \\ 0  & 0 & A_{3} \\
B_{1} & B_{2} & B_{3}\end{array}\!\right]
\!\!\!\!&=&\!\!\!\!
\left[\!\begin{array}{cccc}
0 & I & 0 & 0 \\ I & 0 & 0 & 0 \\ 0  & 0 & I & 0 \\ 0 & 0 & 0 & I \end{array}\!\right] \!
\left[\!\begin{array}{ccc}
A_{2} & 0 & 0 \\ 0 & A_{1} & 0 \\ 0  & 0 & A_{3} \\
B_{2} & B_{1} & B_{3}\end{array}\!\right]\!\! \times \nonumber\\
& & \hspace*{2cm}
\left[\begin{array}{ccc}
0 & I & 0 \\ I & 0 & 0 \\ 0  & 0 & I \end{array}\right]
\end{eqnarray}
in which the identity matrices and the zero matrices in general have different dimensions. Obviously, a necessary and sufficient condition for the matrix
$\left[\begin{array}{c}
{\rm\bf diag}\left\{A_{1},\;A_{2},\; A_{3}\right\} \\ \left[ B_{1} \;\; B_{2} \;\; B_{3}\right]\end{array}\right]$ being of FCR is that the matrix
$\left[\begin{array}{c}
{\rm\bf diag}\left\{A_{2},\;A_{1},\; A_{3}\right\} \\ \left[ B_{2} \;\; B_{1} \;\; B_{3}\right]\end{array}\right]$ has this property.

When the matrix $A_{2}$ is of FCR, direct algebraic operations show that $[A_{2}\; 0\; 0]^{\perp}\!=\!{\rm\bf col}\left\{[0\;\; 0],\; [I\;\; 0],\;[0\;\; I]\right\}$. As
\begin{equation}
\left[\!\begin{array}{ccc}
0 & A_{1} & 0 \\ 0  & 0 & A_{3} \\
B_{2} & B_{1} & B_{3}\end{array}\!\right]\!\!
\left[\begin{array}{cc}
0 & 0 \\ I & 0 \\ 0 & I \end{array}\right]
=
\left[\!\begin{array}{cc}
A_{1} & 0 \\  0 & A_{3} \\
B_{1} & B_{3}\end{array}\!\right]
\end{equation}
the proof can be completed through a direct application of Lemma \ref{lemma:3}.   \hspace{\fill}$\Diamond$

From this lemma, the following conclusions can be directly obtained. As the proof is quite straightforward, it is omitted.
\begin{corollary}
Assume that $A_{i}^{[j]}|_{i=1,j=1}^{i=3,j=m}$ and $B_{i}^{[j]}|_{i=1,j=1}^{i=3,j=m}$ are some matrices having compatible dimensions, and the matrix $\left[A_{2}^{[1]}\;A_{2}^{[2]}\;\cdots\; A_{2}^{[m]}\right]$ is of FCR. Then the matrix
\begin{displaymath}
\left[\!\!\begin{array}{ccc}
{\rm\bf diag}\!\left\{\!A_{1}^{[1]},\;A_{2}^{[1]},\; A_{3}^{[1]}\!\right\} &  \cdots &
{\rm\bf diag}\!\left\{\!A_{1}^{[m]},\;A_{2}^{[m]},\; A_{3}^{[m]}\!\right\}   \\
\left[ B_{1}^{[1]} \;\;\;\; B_{2}^{[1]} \;\;\;\; B_{3}^{[1]}\right] &  \cdots &
\left[ B_{1}^{[m]} \;\;\;\; B_{2}^{[m]} \;\;\;\; B_{3}^{[m]}\right]\end{array}\!\!\right]
\end{displaymath}
is of FCR, if and only if the following matrix has this property
\begin{displaymath}
\left[\begin{array}{ccc}
{\rm\bf diag}\left\{A_{1}^{[1]},\; A_{3}^{[1]}\right\} &  \cdots &
{\rm\bf diag}\left\{A_{1}^{[m]},\; A_{3}^{[m]}\right\}   \\
\left[ B_{1}^{[1]} \;\;\;\;  B_{3}^{[1]}\right] &  \cdots &
\left[ B_{1}^{[m]} \;\;\;\;  B_{3}^{[m]}\right]\end{array}\right]
\end{displaymath}
\label{corollary:2}
\end{corollary}

In order to derive a computationally scalable necessary and sufficient condition for NDS observability/controllability verifications, the following results on matrix pencils are introduced, which are given in many published works including \cite{bv88,it17}.

\begin{definition}
Let $G$ and $H$ be two arbitrary $m\times n$ dimensional real matrices. A  matrix valued polynomial $\Psi(\lambda)=\lambda G+H$ is called a matrix pencil.
\begin{itemize}
\item This matrix pencil is called regular, whenever $m=n$ and ${\rm\bf det}(\Psi(\lambda))\not\equiv 0$.
\item If both the matrices $G$ and $H$ are invertible, then this matrix pencil is called strictly regular.
\item If there exist two nonsingular real matrices $U$ and $V$, such that  $\Psi(\lambda)=U\bar{\Psi}(\lambda)V$ are satisfied by two matrix pencils ${\Psi}(\lambda)$ and $\bar{\Psi}(\lambda)$, then these two matrix pencils are said to be strictly equivalent.
\end{itemize}
\end{definition}

Throughout this paper, for an arbitrary positive integer $m$, the symbol $H_{m}(\lambda)$ stands for an $m\times m$ dimensional strictly regular matrix pencil, while the symbols $K_{m}(\lambda)$, $N_{m}(\lambda)$, $L_{m}(\lambda)$ and $J_{m}(\lambda)$ respectively for matrix pencils having the following definitions,
\begin{eqnarray}
& &
\hspace*{-1.0cm} K_{m}(\lambda)\!=\!\lambda I_{m}\!+\!\left[\!\!\begin{array}{cc}
0 & I_{m-1} \\ 0 & 0 \end{array}\!\!\!\right]\!,\hspace{0.1cm}
N_{m}(\lambda)\!=\!\lambda \!\left[\!\!\begin{array}{cc}
0 & I_{m-1} \\ 0 & 0 \end{array}\!\!\!\right] \!+\! I_{m} \label{eqn:4} \\
& &
\hspace*{-1.0cm} L_{m}(\lambda)=\left[\begin{array}{cc}
K_{m}(\lambda) & \left[\begin{array}{c} 0 \\ 1 \end{array}\right] \end{array}\right],\hspace{0.15cm}
J_{m}(\lambda)= \left[\begin{array}{c}
K_{m}^{T}(\lambda) \\ \left[0 \;\;\;\;\; 1\right] \end{array}\right] \label{eqn:5}
\end{eqnarray}
These matrix pencils are often used in constructing the Kronecker canonical form (KCF) of a general matrix pencil. Obviously, the dimensions of the matrix pencils $K_{m}(\lambda)$ and $N_{m}(\lambda)$ are $m\times m$, while the matrix pencils $L_{m}(\lambda)$ and $J_{m}(\lambda)$ respectively have a dimension of $m\times (m+1)$ and $(m+1)\times m$. Moreover, when $m=0$, $L_{m}(\lambda)$ is a $0\times 1$ zero matrix whose existence means adding a zero column vector in a KCF without increasing its rows, while $J_{m}(\lambda)$ is a $1\times 0$ zero matrix whose existence means adding a zero row vector in a KCF without increasing its columns. On the other hand, $J_{m}(\lambda)=L_{m}^{T}(\lambda)$. For a clear presentation, however, it appears better to introduce these two matrix pencils simultaneously.

In other words, the capital letters $H$, $K$, $N$, $J$ and $L$ are used in this paper to indicate the type of the associated matrix pencil, while the subscript $m$ its dimensions.

When a matrix pencil is block diagonal with the diagonal blocks having the form $H_{*}(\lambda)$, $K_{*}(\lambda)$, $N_{*}(\lambda)$, $L_{*}(\lambda)$ and $J_{*}(\lambda)$, it is called KCF. It is now extensively known that any matrix pencil is strictly equivalent to a KCF \cite{bv88,Gantmacher59,it17}, which can be stated as follows.

\begin{lemma}
For any matrix pencil $\Psi(\lambda)$, there are some unique nonnegative integers $\xi_{\rm\bf H}$, $\zeta_{\rm\bf K}$, $\zeta_{{\rm\bf L}}$, $\zeta_{{\rm\bf N}}$, $\zeta_{{\rm\bf J}}$, $\xi_{\rm\bf L}(j)|_{j=1}^{\zeta_{{\rm\bf L}}}$ and $\xi_{\rm\bf J}(j)|_{j=1}^{\zeta_{{\rm\bf J}}}$, as well as some unique positive integers $\xi_{\rm\bf K}(j)|_{j=1}^{\zeta_{{\rm\bf K}}}$ and $\xi_{\rm\bf N}(j)|_{j=1}^{\zeta_{{\rm\bf N}}}$, such that $\Psi(\lambda)$ is strictly equivalent to the block diagonal matrix pencil $\bar{\Psi}(\lambda)$ defined as
\begin{eqnarray}
\bar{\Psi}(\lambda)\!\!&=&\!\!{\rm\bf diag}\!\left\{\!H_{\xi_{{\rm\bf H}}}(\lambda),\;K_{\xi_{\rm\bf K}(j)}(\lambda)|_{j=1}^{\zeta_{{\rm\bf K}}},\; L_{\xi_{\rm\bf L}(j)}(\lambda)|_{j=1}^{\zeta_{{\rm\bf L}}}, \right.\nonumber \\
& & \hspace*{2.0cm}\left. N_{\xi_{\rm\bf N}(j)}(\lambda)|_{j=1}^{\zeta_{{\rm\bf N}}},\; J_{\xi_{\rm\bf J}(j)}(\lambda)\!|_{j=1}^{\zeta_{{\rm\bf J}}}\!\right\}
\label{eqn:6}
\end{eqnarray}
\label{lemma:2}
\end{lemma}

The following lemma explicitly characterizes the null spaces of the matrix pencils $H_{*}(\lambda)$, $K_{*}(\lambda)$, $N_{*}(\lambda)$, $L_{*}(\lambda)$ and $J_{*}(\lambda)$. This characterization is helpful in clarifying subsystems with which an observable/controllable NDS can be more easily constructed. Its proof is deferred to Appendix A.

\begin{lemma} Let $m$ be an arbitrary positive integer. Then the matrix pencils defined respectively in Equations (\ref{eqn:4}) and (\ref{eqn:5}) have the following null spaces.
\begin{itemize}
\item $H_{m}(\lambda)$ is not of full rank (FR) only at $m$ isolated complex values of the variable $\lambda$. All these values are not equal to zero.
\item $N_{m}(\lambda)$ is always of FR.
\item $J_{m}(\lambda)$ is always of FCR.
\item $K_{m}(\lambda)$ is singular only at $\lambda=0$, and $K_{m}^{\perp}(0)={\rm\bf col}\left\{1, 0_{m-1}\right\}$.
\item $L_{m}(\lambda)$ is not of FCR at every complex $\lambda$, and
$L_{m}^{\perp}(\lambda)={\rm\bf col}\left\{\left. 1, (-\lambda)^{j}\right|_{j=1}^{m}\right\}$.
\end{itemize}
\label{lemma:0}
\end{lemma}

\section{Observability and Controllability of an NDS}

Note that parallel, cascade and feedback connections of LFTs can still be expressed as an LFT \cite{zdg96}. On the other hand, \cite{zhou15} has already made it clear that the system matrices of the whole NDS can be represented as an LFT of its SCM, provided that all the subsystems are connected by their internal inputs/outputs. These make it possible to rewrite the NDS $\bf\Sigma$  in a form which is completely the same as that of \cite{zhou15}, in which all the (pseudo) FPPs of each subsystem, as well as the subsystem connection matrix, are included in a single matrix. This has also been performed in \cite{zz19}. As the associated expressions are important in studying actuator/sensor placements in Section V, and the derivations are not very lengthy, both of them are included in this section to make the presentation more easily understandable.

More specifically, for every subsystem ${\bf{\Sigma}}_i$ with $i\in\{1,2,\cdots,N\}$, the following two auxiliary internal output and input vectors $z^{[a]}(t,i)$ and $v^{[a]}(t,i)$ are introduced,
\begin{eqnarray}
& & \hspace*{-1cm} z^{[a]}(t,i) = [{F_1{(i)}}\  {F_2{(i)}}\  {F_3{(i)}}]\!\left[\!\!\begin{array}{c} x(t,i) \\ v(t,i) \\ u(t,i) \end{array}\!\!\right] \!+\! {G{(i)}}v^{[a]}(t,i)
\label{eqn:7} \\
& & \hspace*{-1cm} v^{[a]}(t,i) = P(i)z^{[a]}(t,i)
\label{eqn:8}
\end{eqnarray}
Define vectors $\bar{z}(t,i)$ and $\bar{v}(t,i)$ respectively as  $\bar{z}(t,i)={\rm\bf col}\{z(t,i),z^{[a]}(t,i)\}$ and $\bar{v}(t,i) ={\rm\bf col}\{v(t,i),v^{[a]}(t,i)\}$, and assume that their dimensions are respectively   $m_{{\rm\bf\bar{z}}i}$ and $m_{{\rm\bf\bar{v}}i}$. Straightforward algebraic manipulations show that when this subsystem is well-posed, i.e, when the matrix $I-G{(i)} P{(i)}$ is regular, its input-output relations can be rewritten as (\ref{eqn:8}) and the next equation
\begin{equation}
\left[ \begin{matrix}
\delta(x(t,i)) \\
{{\bar{z}}(t,i)}\\
{{y}(t,i)}
\end{matrix} \right] = \left[ \begin{matrix}
{A_{\rm\bf xx}{(i)}} & {A_{\rm\bf xv}{(i)}} & {B_{\rm\bf x}{(i)}}\\
{A_{\rm\bf zx}{(i)}} & {A_{\rm\bf zv}{(i)}} & {B_{\rm\bf z}{(i)}}\\
{C_{\rm\bf x}{(i)}} & {C_{\rm\bf v}{(i)}} & {D{(i)}}
\end{matrix} \right]\left[ \begin{matrix}
{x(t,i)}\\
{\bar{v}(t,i)}\\
{u(t,i)}
\end{matrix} \right]
\label{eqn:11}
\end{equation}
in which $D{(i)} = D^{[0]}{(i)}$ and
\begin{eqnarray*}
& & \hspace*{-0.8cm} A_{\rm\bf xx}{(i)}=A_{\rm\bf xx}^{[0]}{(i)},\hspace{0.1cm}  A_{\rm\bf xv}{(i)} = \left[ \begin{matrix}
{A_{\rm\bf xv}^{[0]}{(i)}}& {H_1{(i)}}
\end{matrix} \right] \\
& & \hspace*{-0.8cm}  A_{\rm\bf zx}{(i)}{\rm{ = }}\left[ \begin{matrix}
{A_{\rm\bf zx}^{[0]}{(i)}}  \\
{F_1{(i)}}
\end{matrix} \right],\hspace{0.1cm} A_{\rm\bf zv}{(i)} = \left[ \begin{matrix}
{A_{\rm\bf zv}^{[0]}{(i)}} & {H_2{(i)}}\\
{F_2{(i)}} & {{G{(i)}}}
\end{matrix} \right] \\
 & & \hspace*{-0.8cm} B_{\rm\bf x}^{(i)}=B_{\rm\bf x}^{[0]}{(i)}, \hspace{0.25cm} B_{\rm\bf z}{(i)} = \left[ \begin{matrix}
{B_{\rm\bf z}^{[0]}{(i)}}\\
{F_3{(i)}}
\end{matrix} \right] \\
& & \hspace*{-0.8cm} C_{\rm\bf x}(i) = C_{\rm\bf x}^{[0]}{(i)}, \hspace{0.25cm} C_{\rm\bf v}(i) = \left[\begin{matrix}
{C_{\rm\bf v}^{[0]}{(i)}}&{H_3{(i)}}
\end{matrix} \right]
\end{eqnarray*}

Denote vectors ${\rm\bf col}\!\left\{\! \bar{z}(t,i)|_{i=1}^N\!\right\}$ and ${\rm\bf col}\!\left\{\! \bar{v}(t,i)|_{i=1}^N\!\right\}$ respectively by
$\bar{z}(t)$ and $\bar{v}(t)$. Moreover, construct a matrix $\bar{\Phi}$ as
\begin{equation}
\bar{\Phi}  = \begin{pmat}[{.|.||}]
{{\Phi _{11}}}&{}&{{\Phi _{12}}}&{}&{\cdots}&{{\Phi _{1N}}}&{}\cr
{}&{ P{(1)}}&{}&0& \cdots &{}&0\cr\-
 \vdots & \vdots & \vdots & \vdots & \ddots & \vdots & \vdots \cr\-
{{\Phi _{N1}}}&{}&{{\Phi _{N2}}}&{}&{\cdots}&{{\Phi _{NN}}}&{}\cr
{}&0&{}&0&{\cdots}&{}&{ P{(N)}}\cr
\end{pmat}
\label{eqn:10}
\end{equation}
Then Equations (\ref{eqn:2}) and (\ref{eqn:8}) can be compactly rewritten as
\begin{equation}
\bar{v}(t) =  \bar{\Phi} \bar{z}(t)
\label{eqn:9}
\end{equation}
Here, $\Phi_{ij}$ with $i,j\in\{1,2,\cdots,N\}$ stands for the $(i,j)$-th submatrix of the SCM $\Phi$ when it is divided compatibly with the dimensions of the system internal output and input vectors.

To emphasize similarities in system analyses and syntheses between the matrices $\bar{\Phi}$ and $\Phi$, as well as to distinguish their engineering significance, etc., the matrix $\bar{\Phi}$ is called the augmented SCM in the remaining of this paper.

Equations (\ref{eqn:11}) and (\ref{eqn:9}) give an equivalent description for the input-output relations of the NDS ${\bf\Sigma}$, which has completely the same form as that for the NDS investigated in \cite{zhou15}. This equivalent form is benefited from the invariance properties of LFTs, and makes results of \cite{zhou15} straightforwardly applicable to the NDS $\bf\Sigma$, which are given in the next lemma.

\begin{lemma}
Assume that the NDS ${\rm\bf\Sigma}$, and each of its subsystems ${\bf{\Sigma}}_i|_{i=1}^{N}$ , are well-posed. Then this NDS is observable if and only if for an arbitrary complex scalar $\lambda$, the following matrix pencil $M(\lambda)$ is of FCR,
\begin{equation}
M(\lambda)=\left[\begin{array}{cc}
\lambda I_{M_{\rm\bf x}}-A_{\rm\bf xx} & -A_{\rm\bf xv} \\
-C_{\rm\bf x} & -C_{\rm\bf v} \\
-\bar{\Phi} A_{\rm\bf zx} & I_{M_{\rm\bf z}}-\bar{\Phi} A_{\rm\bf zv} \end{array}\right]
\label{eqn:3}
\end{equation}
\label{lemma:1}
\end{lemma}
Here, $A_{\rm\bf
*\#}\!\!=\!\!{\rm\bf diag}\!\left\{\!A_{\rm\bf
*\#}(i)|_{i=1}^{N}\!\right\}$, $C_{\rm\bf *}\!\!=\!\!{\rm\bf diag}\!\!\left\{\!C_{\rm\bf
*}(i)|_{i=1}^{N}\!\right\}$, in which
${\rm\bf *,\#}={\rm\bf x}$, ${\rm\bf v}$, or ${\rm\bf z}$.

Assume that there are $M$ subsystems in the NDS $\bf\Sigma$, denote their indices by $k(j)|_{j=1}^{M}$, with their system matrix $[C_{\rm\bf x}(k(j))\;\;C_{\rm\bf v}(k(j))]$ not being of FCR. For a clear presentation, assume without any loss of generality that $1\leq k(1)<k(2)<\cdots<k(M)\leq N$. Then from matrix theories \cite{Gantmacher59,hj91}, we have that $[C_{\rm\bf x}(k(j))\;\;C_{\rm\bf v}(k(j))]^{\perp}$ is of FCR for each $j=1,2,\cdots,M$. Let $N_{\rm\bf cx}(k(j))$ and $N_{\rm\bf cv}(k(j))$ denote respectively the first $m_{{\rm\bf x}k(j)}$ rows and the remaining $m_{\bar{\rm\bf v}k(j)}$ rows of this matrix. Using these notations, the following conclusions are derived, which establish some relations between the observability of the NDS investigated in this paper and that of a descriptor system. Their proof is deferred to Appendix B.

\begin{theorem}
Define matrices $N_{\rm\bf cx}$ and $N_{\rm\bf cv}$ respectively as
\begin{eqnarray}
& & N_{\rm\bf cx}\!=\!\!\left[\!\!\begin{array}{c}  0  \\ {\rm\bf diag}\!\left\{\!\!\left.\left[\!\!\begin{array}{c} N_{\rm\bf cx}(k(j)) \\  0 \end{array}\!\!\right]\!\right|_{j=1}^{M}\!\right\}\end{array}\!\!\!\right]\!
\label{eqn:13a}  \\
& & N_{\rm\bf cv}\!=\!\left[\!\!\begin{array}{c} 0 \\ {\rm\bf diag}\!\left\{\!\!\left.\left[\!\!\begin{array}{c} N_{\rm\bf cv}(k(j)) \\ 0 \end{array}\!\!\!\right]\!\right|_{j=1}^{M}\!\!\right\}\end{array}\!\!\!\right]
\label{eqn:13b}
\end{eqnarray}
in which the zero matrices in general have different dimensions. Then the NDS $\bf\Sigma$  is observable if and only if for every complex scalar $\lambda$, the following matrix pencil $\Psi(\lambda)$ is of FCR,
\begin{equation}
\Psi(\lambda)=\lambda\left[\begin{array}{c}
N_{\rm\bf cx} \\  0 \end{array}\right]+
\left[\begin{array}{c}
-A_{\rm\bf xx}N_{\rm\bf cx}-A_{\rm\bf xv}N_{\rm\bf cv} \\
   N_{\rm\bf cv}-\bar{\Phi} \left(A_{\rm\bf zx}N_{\rm\bf cx}+A_{\rm\bf zv}N_{\rm\bf cv}\right) \end{array}\right]
\label{eqn:3-a}
\end{equation}
\label{theorem:1}
\end{theorem}

\begin{remark}
When the matrix $N_{\rm\bf cx}$ is square and the matrix
${\rm\bf col}\!\left\{\!N_{\rm\bf cx},\; N_{\rm\bf cv}\!-\!\bar{\Phi} \left(A_{\rm\bf zx}N_{\rm\bf cx}+A_{\rm\bf zv}N_{\rm\bf cv}\right)\!\right\}$ is of FCR, as well as that ${\rm\bf det} \left(\lambda N_{\rm\bf cx} -\left[ A_{\rm\bf xx}N_{\rm\bf cx}+A_{\rm\bf xv}N_{\rm\bf cv}\right] \right)\not\equiv 0$, the condition that the aforemention matrix pencil $\Psi(\lambda)$ is of FCR at every complex $\lambda$, is necessary and sufficient for the observability of the following descriptor system,
\begin{eqnarray*}
& & N_{\rm\bf cx}\delta(x(t))= \left[A_{\rm\bf xx}N_{\rm\bf cx}+A_{\rm\bf xv}N_{\rm\bf cv}\right]x(t) \\
& & y(t)=\left[ N_{\rm\bf cv}-\bar{\Phi} \left(A_{\rm\bf zx}N_{\rm\bf cx}+A_{\rm\bf zv}N_{\rm\bf cv}\right) \right] x(t)
\end{eqnarray*}
Results on the observability of descriptor systems appear directly applicable to that of the NDS $\bf\Sigma$, while the former has been extensively studied and various conclusions have been established \cite{Dai89,Duan10}. A direct application of these results, however, can not efficiently use the block diagonal structure of the associated matrices, and usually introduces some unnecessary computational costs that is not quite attractive for large scale NDS analysis and synthesis. In addition, there are in general not any guarantees that the matrix pencil $\lambda N_{\rm\bf cx} -\left[ A_{\rm\bf xx}N_{\rm\bf cx}+A_{\rm\bf xv}N_{\rm\bf cv}\right]$ is regular. As a matter of fact, this matrix pencil may sometimes even not be square. Furthermore, the matrix
${\rm\bf col}\!\left\{\!N_{\rm\bf cx},\; N_{\rm\bf cv}\!-\!\bar{\Phi} \left(A_{\rm\bf zx}N_{\rm\bf cx}+A_{\rm\bf zv}N_{\rm\bf cv}\right)\!\right\}$ depends on the augmented SCM $\bar{\Phi}$, and can not be guaranteed to be always of FCR.
\end{remark}

\begin{remark}
The results of Theorem \ref{theorem:1} essentially mean that when there is a subsystem in the NDS $\bf\Sigma$ with its external output matrix, that is, $[C_{\rm\bf x}(i)\;\;C_{\rm\bf v}(i)]$, being of FCR, then the conclusions of system observability are not changed by removing the column blocks associated with this subsystem from the matrix pencil $M(\lambda)$. This can also be understood from Corollary \ref{corollary:2}. This property is quite interesting in NDS designs, as it means that if a subsystem holds this property, then it will not affect the observability of the whole NDS, no matter how it is  connected to other subsystems.
\end{remark}

To derive a computationally attractive condition for the observability of the NDS of Equations (\ref{eqn:1}) and (\ref{eqn:2}), the KCF for a matrix pencil, which is given in the previous section, are helpful.

It can be claimed from Lemma \ref{lemma:2} that for each $i\in\{k(1),k(2),\cdots,k(M)\}$, there are regular real matrices $U(i)$ and $V(i)$, and a matrix pencil $\Xi(\lambda,i)$, such that
\begin{equation}
\lambda N_{\rm\bf cx}(i)-\left[A_{\rm\bf xx}(i)N_{\rm\bf cx}(i)+A_{\rm\bf xv}(i)N_{\rm\bf cv}(i)\right]=U(i)\Xi(\lambda,i)V(i)
\label{eqn:14}
\end{equation}
in which
\begin{eqnarray}
{\Xi}(\lambda,i)\!\! & =& \!\!{\rm\bf diag}\!\left\{\!H_{\xi_{{\rm\bf H}i}}(\lambda),\;K_{\xi_{{\rm\bf K}i}(j)}(\lambda)|_{j=1}^{\zeta_{{\rm\bf K}i}},\; L_{\xi_{{\rm\bf L}i}(j)}(\lambda)|_{j=1}^{\zeta_{{\rm\bf L}i}}, \;  \right.\nonumber\\
& & \hspace*{1.5cm} \left. N_{\xi_{{\rm\bf N}i}(j)}(\lambda)|_{j=1}^{\zeta_{{\rm\bf N}i}},\; J_{\xi_{{\rm\bf J}i}(j)}(\lambda)\!|_{j=1}^{\zeta_{{\rm\bf J}i}}\!\right\}
\label{eqn:15}
\end{eqnarray}
Here, $\xi_{{\rm\bf H}i}$, $\zeta_{{\rm\bf K}i}$, $\zeta_{{\rm\bf N}i}$, $\zeta_{{\rm\bf L}i}$,  $\zeta_{{\rm\bf J}i}$, $\xi_{{\rm\bf L}i}(j)|_{j=1}^{\zeta_{{\rm\bf L}i}}$ and
$\xi_{{\rm\bf J}i}(j)|_{j=1}^{\zeta_{{\rm\bf J}i}}$ are some nonnegative integers, $\xi_{{\rm\bf K}i}(j)|_{j=1}^{\zeta_{{\rm\bf K}i}}$ and $\xi_{{\rm\bf N}i}(j)|_{j=1}^{\zeta_{{\rm\bf N}i}}$ are some positive integers. All these numbers are uniquely determined by the system matrices $A_{\rm\bf xx}(i)$, $A_{\rm\bf xv}(i)$, $N_{\rm\bf cx}(i)$ and $N_{\rm\bf cv}(i)$ of the $i$-th subsystem ${\rm\bf\Sigma}_{i}$.

In the decomposition of Equation (\ref{eqn:14}), the calculations are performed for each subsystem individually. On the other hand, there are extensive studies on expressing a matrix pencil with the KCF and various computationally attractive algorithms have already been established \cite{bv88,Gantmacher59}. It can therefore be declared that computations involved in the aforementioned decomposition are in general possible, while the total computational complexity increases linearly with the increment of the subsystem number $N$.

Define matrix pencils $\bar{\Xi}(\lambda)$ and $\bar{\Xi}(\lambda,i)$ with $i\in\{k(1),k(2),\cdots,k(M)\}$ respectively as
\begin{eqnarray*}
& & \hspace*{-0.5cm} \bar{\Xi}(\lambda,i)={\rm\bf diag}\!\left\{\!H_{\xi_{{\rm\bf H}i}}(\lambda),\,K_{\xi_{{\rm\bf K}i}(j)}(\lambda)|_{j=1}^{\zeta_{{\rm\bf K}i}},\, L_{\xi_{{\rm\bf L}i}(j)}(\lambda)|_{j=1}^{\zeta_{{\rm\bf L}i}}\!\!\right\}  \\
& & \hspace*{-0.5cm} \bar{\Xi}(\lambda)={\rm\bf diag}\!\left\{\!\bar{\Xi}(\lambda,k(i))|_{i=1}^{M}\right\}
\end{eqnarray*}
Based on the above observations, as well as the expressions in Equation (\ref{eqn:15}), the following condition is obtained for NDS observability. Its derivations are deferred to Appendix C.

\begin{theorem}
For every $i\in\{k(1),k(2),\cdots,k(M)\}$, denote $\xi_{{\rm\bf H}i}+\sum_{j=1}^{\zeta_{{\rm\bf K}i}}\xi_{{\rm\bf K}i}(j)+\sum_{j=1}^{\zeta_{{\rm\bf L}i}}\xi_{{\rm\bf L}i}(j)$ by $s(i)$. Moreover, represent the matrix constructed from the first $s(i)$ columns of the inverse of the matrix $V(i)$ by $V_{i}^{-1}(1:s(i))$. Denote the following matrix pencil
\begin{displaymath}
\hspace*{-0.2cm}\left[\!\!\!\!\begin{array}{c}
\bar{\Xi}(\lambda) \\
\left[\! N_{\rm\bf cv}-\bar{\Phi}(A_{\rm\bf zx}N_{\rm\bf cx}+A_{\rm\bf zv}N_{\rm\bf cv})\!\right]\!{\rm\bf diag}\!\left\{\! V_{k(i)}^{-1}(1:s(k(i)))|_{i=1}^{M}\!\right\} \end{array}\!\!\!\!\right]
\end{displaymath}
by $\bar{\Psi}(\lambda)$. Then the NDS $\bf\Sigma$  is observable, if and only if the matrix pencil $\bar{\Psi}(\lambda)$ is of FCR at each $\lambda\in {\cal C}$.
\label{theorem:2}
\end{theorem}

From the proof of Theorem \ref{theorem:2}, it is clear that if $\xi_{{\rm\bf H}i}=\zeta_{{\rm\bf K}i}=\zeta_{{\rm\bf L}i}=0$ in each $\Xi(\lambda,i)$ with $i\in\{k(1),k(2),\cdots,k(M)\}$, which is essentially a condition required for each subsystem individually, then the NDS $\bf\Sigma$  is always observable, no matter how the subsystems are connected and the parameters of a subsystem are selected.

For each $i\in\{k(1),k(2),\cdots,k(M)\}$, let ${\rm\bf\Lambda}(i)$ denote the set consisting of the complex $\lambda$ at which $\bar{\Xi}(\lambda,i)$ is not of FCR. From Lemma \ref{lemma:0}, this set is the whole complex plane if $\zeta_{{\rm\bf L}i}\neq 0$. On the other hand, if $\zeta_{{\rm\bf L}i}=0$, then this set is simply formed by zero and all the complex values that lead to a singular $H_{\xi_{{\rm\bf H}i}}(\lambda)$. In addition, $\bar{\Xi}^{\perp}(\lambda,i)$ is also block diagonal with each of its blocks being completely determined by $H^{\perp}_{\xi_{{\rm\bf H}i}}(\lambda)$,   $K^{\perp}_{\xi_{{\rm\bf K}i}(j)}(\lambda)$ and $L^{\perp}_{\xi_{{\rm\bf L}i}(j)}(\lambda)$. Moreover, all of them can be easily obtained using the results of Lemma \ref{lemma:0}.

Furthermore, let ${\rm\bf\Lambda}$ denote the set consisting of the values of the complex variable $\lambda$ at which $\bar{\Xi}(\lambda)$ is not of FCR, and ${\cal M}(\lambda_{0})$ the set of all the subsystem indices with which the matrix pencil $\bar{\Xi}(\lambda,i)$ is not of FCR at $\lambda_{0}$. From the definitions of $\bar{\Xi}(\lambda)$ and $\bar{\Xi}(\lambda,i)|_{i=1}^{N}$, it is obvious that
\begin{equation}
{\rm\bf\Lambda}=\bigcup_{i=1}^{M}{\rm\bf\Lambda}(k(i)), \hspace{0.25cm}
\bar{\Xi}^{\perp}(\lambda_{0})\!=\!\left[\!\!\begin{array}{c} 0 \\ {\rm\bf diag}\!\left\{\!\!\left.
{\left[\!\!\begin{array}{c} \bar{\Xi}^{\perp}(\lambda_{0},i) \\ 0 \end{array}\!\!\right]}\right|_{i\in {\cal M}(\lambda_{0})} \!\right\} \end{array}\!\!\!\right]
\label{eqn:31}
\end{equation}

The following results can be immediately established from Theorem \ref{theorem:2} and Lemma \ref{lemma:3}.

\begin{theorem}
For a prescribed complex $\lambda_{0}$, define matrices $X(\lambda_{0})$ and $Y(\lambda_{0})$ respectively as
\begin{eqnarray}
& &\hspace*{-0.5cm} X(\lambda_{0})\!=\!{\rm\bf diag}\!\left\{\! N_{\rm\bf cv}(k(i))V_{k(i)}^{-1}(1:s(k(i)))|_{i=1}^{M}\!\right\}\bar{\Xi}^{\perp}(\lambda_{0}) \\
& &\hspace*{-0.5cm}Y(\lambda_{0})\!=\!{\rm\bf diag}\!\left\{\!\left[ A_{\rm\bf zx}(k(i))N_{\rm\bf cx}(k(i))\!+\!A_{\rm\bf zv}(k(i)) N_{\rm\bf cv}(k(i))\right]\!\times\right. \nonumber\\
& & \hspace*{2.6cm} \left. V_{k(i)}^{-1}\!(1\!:\!s(k(i)))|_{i\!=\!1}^{M}\!\right\}\!\!\bar{\Xi}^{\perp}(\lambda_{0})
\end{eqnarray}
Then the NDS $\bf\Sigma$  is observable, if and only if for each $\lambda_{0}\in {\rm\bf\Lambda}$, the matrix
\begin{equation}
X(\lambda_{0})-\bar{\Phi} Y(\lambda_{0})
\label{eqn:32}
\end{equation}
is of FCR.
\label{theorem:3}
\end{theorem}

The proof is omitted due to its obviousness.

The above theorem makes it clear that in the matrix pencil $\bar{\Xi}(\lambda)$, the existence of a matrix pencil with  the form of $L_{*}(\lambda)$ may greatly increase difficulties for the satisfaction of the observability requirement by the NDS $\rm\bf\Sigma$, as it makes the set ${\rm\bf\Lambda}$ equal to the whole complex plane and requires that the matrix given by Equation (\ref{eqn:32}) is of FCR at each complex $\lambda_{0}$. The latter leads in general to infinitely many constraints on the augmented SCM $\bar{\Phi}$. It is interesting to see possibilities to avoid occurrence of this type of matrix pencils in subsystem constructions for an NDS. This will be discussed briefly in the following Section IV.

\begin{remark}
In both the definition of the matrix $X(\lambda_{0})$ and the definition of the matrix $Y(\lambda_{0})$, all the involved matrices have a consistent block diagonal structure. This means that these two matrices are also block diagonal, and the computational costs for obtaining them increase only linearly with the increment of the subsystem number $N$. This is a quite attractive property in dealing with a large scale NDS which consists of numerous subsystems.
\end{remark}

\begin{remark}
Note that in the augmented SCM $\bar{\Phi}$, which is defined by Equation (\ref{eqn:10}), both subsystem PFPPs and connection parameters are included. This implies that the above theorem reflects influences of these two kinds of parameters on NDS observability. On the other hand, this augmented SCM $\bar{\Phi}$ clearly has a sparse structure. This means that results about sparse computations, which have been extensively and well studied in fields like numerical analysis, can be applied to the verification of the condition in Theorem \ref{theorem:3}. It is interesting to see possibilities of developing more numerically efficient methods for this condition verification, using the particular sparse structure of the augmented SCM $\bar{\Phi}$ and the consistent block diagonal structure of the matrices $X(\lambda_{0})$ and $Y(\lambda_{0})$.
\end{remark}

\begin{remark}
While the matrices $X(\lambda_{0})$ and $Y(\lambda_{0})$ of Equation (\ref{eqn:32}) are calculated in a significantly different way from those of the previous works reported in \cite{zhou15,zyl18}, they have completely the same form. This implies that the associated NDS observability conditions share the same computational advantages in large scale NDS analyses and syntheses. On the other hand, in the derivations of Theorem \ref{theorem:3}, except the well-posedness assumptions, there are not any other requirements on a subsystem of the NDS $\bf\Sigma$. That is, the FNCR condition on each subsystem, which is required in \cite{zhou15,zyl18} to get the associated transmission zeros of each subsystem, is completely removed. It is worthwhile to mention that there are various actual systems that do not have  this FNCR property. Obvious examples include an NDS with a subsystem that does not have an external output \cite{zyl18}. Removal of this condition is interesting from not only a mathematical viewpoint, but also an application viewpoint.
\end{remark}

\begin{remark}
Compared with \cite{zz19}, the results of the above theorem are in a pure algebraic form. In system analysis and synthesis, they are not as illustrative as the results of \cite{zz19} which are given in a graphic form. It is interesting to see whether or not a graphic form can be obtained from Theorem \ref{theorem:3} on the observability of an NDS. On the other hand, \cite{zz19} proves that structural observability verification is in general NP hard for the NDS $\rm\bf\Sigma$, and computationally feasible results are derived only for the case in which the augmented SCM $\bar{\Phi}$ is diagonal. This requirement can not be easily satisfied by a practical system and significantly restricts their applicability. In the derivations of Theorem \ref{theorem:3}, however, except well-posedness of each subsystem and the whole system, which is also asked in \cite{zz19} and is necessary for a system to properly work, there are not any other constraints on either a subsystem or the whole system of the NDS. This is a significant advantage of the results in this paper over those of \cite{zz19}, and makes controllability/observability verification for a large scale NDS much more computationally feasible than their structural counterparts. 
\end{remark}

\begin{remark}
Recall that controllability of an LTI system is equal to observability of its dual system, and this is also true for an NDS \cite{zhou15,zyl18}. This means that the above results can be directly applied to controllability analysis for the NDS $\bf\Sigma$. As a matter of fact, using the duality between controllability and observability of a system, as well as the equivalence representation of the NDS $\bf\Sigma$, which is given by Equations (\ref{eqn:11}) and (\ref{eqn:9}), it can be declared that the NDS $\bf\Sigma$  is controllable, if and only if the following matrix pencil is of FRR for each complex $\lambda$,
\begin{displaymath}
\left[\begin{array}{ccc}
\lambda I_{M_{\rm\bf x}} - A_{\rm\bf xx} & B_{\rm\bf x}  & -A_{\rm\bf xv}\bar{\Phi} \\
-A_{\rm\bf zx} & B_{\rm\bf z} &    I_{M_{\rm\bf z}} - A_{\rm\bf zv}\bar{\Phi}
\end{array}\right]
\end{displaymath}
in which $B_{\rm\bf x}\!\!=\!\!{\rm\bf diag}\!\left\{\!B_{\rm\bf x}(i)|_{i=1}^{N}\!\right\}$, $B_{\rm\bf z}\!\!=\!\!{\rm\bf diag}\!\left\{\!B_{\rm\bf z}(i)|_{i=1}^{N}\!\right\}$, and all the other matrices have the same definitions as those of Lemma \ref{lemma:1}. Using the KCF of a matrix pencil, as well as a basis for the left null space of the matrix ${\rm\bf col}\!\left\{\! B_{\rm\bf x},\; B_{\rm\bf z} \!\right\}$, similar algebraic manipulations give a necessary and sufficient condition for NDS controllability with a similar form as that of Theorem \ref{theorem:3}.
\end{remark}

\section{Conditions for Sensor/Actuator Placements}

Sections III makes it clear that for any $1\leq i\leq N$, the existence of a matrix pencil $L_{*}(\lambda)$/$J_{*}(\lambda)$ in the matrix pencil $\bar{\Xi}(\lambda,i)$, which is completely and independently determined by the system matrices of the subsystem ${\rm\bf\Sigma}_{i}$, may make the observability/controllability condition difficult to be satisfied by an NDS. An interesting issue is therefore that in constructing subsystems of an NDS, whether it is possible to avoid the existence of this type of matrix pencils. In this section, we investigate how to avoid this occurrence under the assumption that $C_{\rm\bf v}(i)=0$ for every $i=1,2,\cdots, N$.

Note that in the matrix pencil $\bar{\Xi}(\lambda,i)$, both the matrices $C_{\rm\bf x}(i)$ and $C_{\rm\bf v}(i)$ are involved which respectively associate the external output vector $y(t,i)$ of the subsystem ${\rm\bf \Sigma}_{i}$ to its state vector and internal input vector. On the other hand, sensors are usually used for state measurement in a system, and an essential requirement for sensor placements is that the associated system is observable. These mean that the aforementioned issue is closely related to NDS sensor placements which is also an important topic in system designs. In addition, $y(t,i)$ may also be used to evaluate the performances of a system in its designs. Examples include to ask some states of a subsystem to track an objective signal, etc. In either of these situations, the external output vector $y(t,i)$ usually include only some states of a subsystem \cite{zdg96,zyl18}. These observations mean that it does not introduce very severe restrictions in actual applications through assuming that $C_{\rm\bf v}^{[0]}(i)=0$ and $H_{3}(i)=0$ for each $i=1,2,\cdots,N$ in the NDS $\rm\bf\Sigma$. Under this hypothesis, it is obvious from its definition that $C_{\rm\bf v}(i)=0$. Hence, the assumption adopted in this section is reasonable and not quite restrictive.

To settle the above issue, the following results are helpful which are standard in the analysis of a matrix pencil \cite{Gantmacher59,it17}.

\begin{lemma}
Define the rank of an $m\times n$ dimensional matrix pencil $\Psi(\lambda)$ as the maximum dimension of its submatrix whose determinant is not constantly equal to zero, and denote it by ${\rm\bf rank}\left\{\Psi(\lambda)\right\}$. Assume that in the KCF of the matrix pencil $\Psi(\lambda)$, there are $p$ matrix pencils in the form of $L_{*}(\lambda)$ and $q$ matrix pencils in the form of $J_{*}(\lambda)$. Then
\begin{equation}
p=n-{\rm\bf rank}\left\{\Psi(\lambda)\right\},\hspace{0.5cm}
q=m-{\rm\bf rank}\left\{\Psi(\lambda)\right\}
\label{eqn:36}
\end{equation}
\label{lemma:4}
\end{lemma}

An immediate result of Lemma \ref{lemma:4} is that there does not exist a $L_{*}(\lambda)$/$J_{*}(\lambda)$ in the KCF of a matrix pencil, if and only if it is of FNCR/FNRR.

\begin{remark}
From the definition of the matrix pencil $\bar{\Xi}(\lambda,i)$, it is obvious that the nonexistence of a matrix pencil with the form of $L_{*}(\lambda)$ is equal to that in the KCF ${\Xi}(\lambda,i)$. On the other hand, Lemmas \ref{lemma:3} and \ref{lemma:4}, together with Equation (\ref{eqn:14}), imply that in the KCF ${\Xi}(\lambda,i)$, $\zeta_{{\rm\bf L}i}=0$ if and only if the following matrix pencil is of FNCR,
\begin{displaymath}
\left[\begin{array}{cc}
\lambda I_{m_{{\rm\bf x}i}}-A_{\rm\bf xx}(i) & -A_{\rm\bf xv}(i) \\
-C_{\rm\bf x}(i) & -C_{\rm\bf v}(i)  \end{array}\right]
\end{displaymath}
It can therefore be declared from Lemma \ref{lemma:6} and the consistent block diagonal structure of the matrices $A_{\rm\bf xx}$, $A_{\rm\bf xv}$, $C_{\rm\bf x}$ and $C_{\rm\bf v}$ that, there does not exist a subsystem ${\rm\bf\Sigma}_{i}$ with $1\leq i\leq N$ in the NDS ${\rm\bf\Sigma}$, such that there is a matrix pencil in the form of $L_{*}(\lambda)$ in its associated matrix pencil $\bar{\Xi}(\lambda,i)$, if and only if the following matrix pencil is of FNCR,
\begin{equation}
\left[\begin{array}{cc}
\lambda I_{M_{\rm\bf x}}-A_{\rm\bf xx} & -A_{\rm\bf xv} \\
-C_{\rm\bf x} & -C_{\rm\bf v} \end{array}\right]
\label{eqn:54}
\end{equation}
\end{remark}

\begin{remark}
When $\lambda$ is not equal to an eigenvalue of the matrix $A_{\rm\bf xx}$, it is straightforward to prove that
\begin{displaymath}
\left[ \lambda I_{M_{\rm\bf x}} \!\!-\! A_{\rm\bf xx} \;\;  -\!\!A_{\rm\bf xv} \right]^{\perp}
={\rm\bf col}\left\{\!(\lambda I_{M_{\rm\bf x}}\!\!-\!A_{\rm\bf xx})^{-1}A_{\rm\bf xv},\; I \!\right\}
\end{displaymath}
It can therefore be claimed from Lemma \ref{lemma:3} that the matrix pencil given by Equation (\ref{eqn:54}) is of FNCR, if and only if the transfer function matrix (TFM) $C_{\rm\bf v}+C_{\rm\bf x}(\lambda I_{M_{\rm\bf x}}-A_{\rm\bf xx})^{-1}A_{\rm\bf xv}$ is. Note that this TFM is exactly the TFM $G^{[1]}(\lambda)$ defined in \cite{zhou15,zyl18}. The above discussions mean that while the results of \cite{zhou15,zyl18} are valid only when the TFM $G^{[1]}(\lambda)$ is of FNCR, which appears to be a severe restriction on the applicability of the obtained results, its satisfaction by the subsystems of an NDS may greatly reduce difficulties in constructing an observable NDS. Similar conclusions can also be reached on the corresponding FNRR assumption adopted in \cite{zhou15,zyl18} for controllability verification of an NDS.
\end{remark}

In order to get conditions for the non-existence of a matrix pencil in $\bar{\Xi}(\lambda,i)$ that has the form of $L_{*}(\lambda)$, we at first investigate influences of the rank of the matrix $\left[C_{\rm\bf x}\;\; C_{\rm\bf v}\right]$ on the observability of the NDS $\bf\Sigma$. Standard algebraic manipulations show that removing a row in this matrix that depend linearly on the other rows does not affect observability of the NDS. This conclusion is obvious from an application view of point. To be more specific, a linear dependence of the rows of the matrix $[C_{\rm\bf x}\; C_{\rm\bf v}]$ means that some of the NDS external outputs can be expressed as a linear combination of its other external outputs. Hence, these external outputs do not contain any new information about the NDS states, and their elimination does not have any influences on the observability of the NDS.

These observations, together with Lemma \ref{lemma:1}, further mean that in the investigation of the observability of the NDS $\rm\bf\Sigma$, the assumption will not introduce any lose of generality that the matrix $\left[C_{\rm\bf x}\;\; C_{\rm\bf v}\right]$ is of FRR. From the compatible  block diagonal structures of the matrices $C_{\rm\bf v}$ and $C_{\rm\bf x}$, this hypothesis is obviously equivalent to that the matrix $\left[C_{\rm\bf x}(i)\;\; C_{\rm\bf v}(i)\right]$ is of FRR for each $i=1,2,\cdots,N$.

Under these assumptions, the following conclusions are obtained from Lemma \ref{lemma:4} on the number of matrix pencils $L_{*}(\lambda)$ in the matrix pencil $\Xi(\lambda,i)$ defined by Equations (\ref{eqn:14}) and (\ref{eqn:15}).

\begin{theorem}
Assume that $C_{\rm\bf v}(i)=0$. Define a matrix pencil $\Theta(\lambda,i)$ as
\begin{eqnarray}
\Theta(\lambda,i)\!\!\!\!&=&\!\!\!\!\left\{I-A_{\rm\bf xv}(i)\left[A_{\rm\bf xv}^{T}(i)A_{\rm\bf xv}(i)\right]^{-1}A_{\rm\bf xv}^{T}(i)\right\}\times    \nonumber\\
& & \hspace*{2cm} \left[\lambda I-A_{\rm\bf xx}(i)\right]C_{\rm\bf x}^{\perp}(i)
\label{eqn:37}
\end{eqnarray}
Then there does not exist a matrix pencil in the KCF $\Xi(\lambda,i)$ that has the form $L_{*}(\lambda)$, if and only if
\begin{itemize}
\item the matrix pencil $\Theta(\lambda,i)$ is of FNCR;
\item the matrix $A_{\rm\bf xv}(i)$ is of FCR.
\end{itemize}
\label{theorem:4}
\end{theorem}

A proof of this theorem is provided in Appendix D.

Note that when a matrix is of FCR, its product with an arbitrary nonzero vector of a compatible dimension is certainly not equal to zero. On the other hand, recall that in Subsystem ${\rm\bf \Sigma}_{i}$, the matrix $A_{\rm\bf xv}(i)$ is actually an input matrix connecting its internal inputs, that is, signals sent from other subsystems, to its state vector. The requirement that this matrix is of FCR means that in order to construct an observable NDS, it is appreciative in subsystem selections to guarantee that any nonzero signals received from other subsystems have some influences on a subsystem state. In other words, influences on the states of a subsystem from a signal sent by another subsystem, are not allowed to be killed by any other signal(s) sent by any other subsystem(s).

Generally speaking, the matrix $A_{\rm\bf xv}^{T}(i)A_{\rm\bf xv}(i)$ may not be invertible, which leads to some difficulties in the definition of the matrix pencil $\Theta(\lambda,i)$. However, this matrix is certainly positive definite and therefore has an inverse, provided that the matrix $A_{\rm\bf xv}(i)$ is of FCR. This means that when the second condition in the above theorem is satisfied, the matrix pencil $\Theta(\lambda,i)$ is well defined.

When a problem of sensor placements is under investigation, it is a general situation that one sensor measures only one state of plant. This means that in each row of the matrix $C_{\rm\bf x}(i)$, there is only one nonzero element. On the other hand, previous discussions reveal that the matrix $C_{\rm\bf x}(i)$ can be assumed to be of FRR without any loss of generality. These observations mean that the assumption that each element of the external output vector $y(t,i)$ is related only to one of the states of the subsystem ${\rm\bf\Sigma}_{i}$, as well as the assumption that all the elements of the external output vector $y(t,i)$ are different from each other, generally do not introduce any restrictions in an investigation about sensor placement.

In the subsystem ${\rm\bf\Sigma}_{i}$, denote by
\begin{displaymath}
\left\{\left.k(j,i)\right|j\in\{1,2,\cdots,m_{{\rm\bf y}i}\},\; k(j,i)\in \{1,2,\cdots,m_{{\rm\bf x}i}\}\right\}
\end{displaymath}
the set consisting of its states that are measured by a sensor. The above discussions mean that it can be assumed that $k(j_{1},i)\neq k(j_{2},i)$ whenever $j_{1}\neq j_{2}$ without any loss of generality. It can also be directly shown that the following two assumptions do not introduce any restrictions on sensor placement studies also,
\begin{itemize}
\item $1 \leq k(1,i) < k(2,i) < \cdots < k(m_{{\rm\bf y}i},i) \leq m_{{\rm\bf x}i}$;
\item the $j$-th element of the external output vector $y(t,i)$ is equal to the $k(j,i)$-th element of the state vector $x(t,i)$ of the subsystem ${\rm\bf\Sigma}_{i}$, $j=1,2,\cdots, m_{{\rm\bf y}i}$.
\end{itemize}

The details are omitted due to their straightforwardness. On the other hand, the rationality of these assumptions can be easily understood from an application view of points. In particular, any position rearrangements of the NDS external outputs do not change the total information contained in these outputs about its states. Hence, it is not out of imaginations that these assumptions do not introduce any restrictions on observability analysis for the NDS $\rm\bf\Sigma$.

Let $e_{j,i}^{[y]}$ with $1\leq j\leq m_{{\rm\bf y}i}$ denote the $j$-th canonical basis vector of the Euclidean space ${\cal R}^{m_{{\rm\bf y}i}}$, and $O(j,i)$ the $m_{{\rm\bf y}i} \times \left[ k(j,i)-k(j-1,i)-1\right]$ dimensional zero matrix in which $j=1,2,\cdots, m_{{\rm\bf y}i}+1$ with $k(0,i)$ and $k(m_{{\rm\bf y}i}+1,i)$ being respectively defined as $k(0,i)=0$ and $k(m_{{\rm\bf y}i}+1,i)=m_{{\rm\bf x}i}$. The above discussions show that in a sensor placement problem, it can be generally assumed, without any loss of generality, that
\begin{eqnarray}
C_{\rm\bf x}(i)\!\!\!\!\!&=&\!\!\!\!\!\left[
O(1,i)\;\; e_{1,i}^{[y]} \;\; O(2,i) \;\; e_{2,i}^{[y]} \right. \nonumber\\
& & \hspace*{0.20cm} \left. \cdots \;\; O(m_{{\rm\bf y}i},i) \;\;
e_{m_{{\rm\bf y}i},i}^{[y]}\;\; O(m_{{\rm\bf y}i}+1,i) \right]
\label{eqn:45}
\end{eqnarray}

Under this assumption, the following results can be directly obtained from Theorem \ref{theorem:4}.

\begin{corollary}
Assume that $C_{\rm\bf v}(i)=0$ and $C_{\rm\bf x}(i)$ satisfies Equation (\ref{eqn:45}). Denote the set
\begin{displaymath}
\left\{1,2,\cdots,m_{{\rm\bf x}i}\right\}\backslash \left\{k(1,i), k(2,i), \cdots, k(m_{{\rm\bf y}i},i)\right\}
\end{displaymath}
by ${\cal S}(i)$, and the following matrix pencil
\begin{displaymath}
\left\{\!I\!-\!A_{\rm\bf xv}(i)\left[A_{\rm\bf xv}^{T}(i)A_{\rm\bf xv}(i)\right]^{-1}\!\!A_{\rm\bf xv}^{T}(i)\!\right\}\!\left[\lambda I \!-\! A_{\rm\bf xx}(i)\right]\!I_{m_{{\rm\bf x}i}}({\cal S}(i))
\end{displaymath}
by $\bar{\Theta}(\lambda,i)$. Then there does not exist a matrix pencil in the KCF $\Xi(\lambda,i)$ that has the form $L_{*}(\lambda)$, if and only if
\begin{itemize}
\item the matrix $A_{\rm\bf xv}(i)$ is of FCR.
\item the matrix pencil $\bar{\Theta}(\lambda,i)$ is of FNCR.
\end{itemize}
\label{corollary:1}
\end{corollary}

\hspace*{-0.4cm}{\it\bf Proof:} When Equation (\ref{eqn:45}) is satisfied by the matrix $C_{\rm\bf x}(i)$, direct matrix manipulations show that
\begin{displaymath}
C_{\rm\bf x}^{\perp}(i)={\rm\bf diag}\left\{
\left.\left[\begin{array}{c} I_{k(j,i)-k(j-1,i)-1} \\ 0 \end{array}\right]\right|_{j=1}^{m_{{\rm\bf y}i}},\; I_{m_{{\rm\bf x}i}-k(m_{{\rm\bf y}i},i)}\right\}
\end{displaymath}
Obviously, this matrix can be equivalently expressed as
\begin{displaymath}
C_{\rm\bf x}^{\perp}(i)=\left[ e_{j,i}^{[x]}: \;\; j\in {\cal S}(i) \right]
\end{displaymath}
in which $e_{j,i}^{[x]}$ with $1\leq j\leq m_{{\rm\bf x}i}$ denote the $j$-th canonical basis vector of the Euclidean space ${\cal R}^{m_{{\rm\bf x}i}}$.

The results can then be obtained by applying this expression for the matrix $C_{\rm\bf x}^{\perp}(i)$ to the definition of the matrix pencil $\Theta(\lambda,i)$ given in Theorem \ref{theorem:4}. This completes the proof.   \hspace{\fill}$\Diamond$

From the definitions of the matrices $A_{\rm\bf xx}(i)$ and $A_{\rm\bf xv}(i)$, which are given immediately after Equation (\ref{eqn:11}), it is clear that these two matrices depend neither on a PFPP of a subsystem, nor on a nonzero entry in the SCM. In other words, these two matrices depend only on the principles of mechanics, electricity, chemistry, biology, etc., that govern the dynamics of the subsystem ${\rm\bf\Sigma}_{i}$. Therefore, the results of Theorem \ref{theorem:4} and Corollary \ref{corollary:1} appear very helpful in determining the type of subsystems with which an NDS can be constructed that is possible to achieve good performances more easily.

On the other hand, although the 2nd condition of Corollary \ref{corollary:1} is combinatorial, it depends only on a single subsystem. As the dimension of the state vector in a subsystem is usually not very large, this condition does not lead to a heavy computational burden in general. In addition, using the KCF of the matrix pencil $\left\{\!I\!-\!A_{\rm\bf xv}(i)\left[A_{\rm\bf xv}^{T}(i)A_{\rm\bf xv}(i)\right]^{-1}\!\!A_{\rm\bf xv}^{T}(i)\!\right\}\!\left[\lambda I \!-\! A_{\rm\bf xx}(i)\right]$, some more explicit conditions on sensor positions can be obtained. The details are omitted due to space considerations.

On the basis of the duality between controllability and observability of the NDS $\rm\bf\Sigma$, similar results can be obtained for actuator placements. Details are omitted due to their straightforwardness.

\section{Extensions to an NDS with Descriptor Subsystems}

For various traditional control plants, rather than a state space model, it is more convenient to describe its dynamics by a descriptor form. Typical examples include constrained mechanical systems, electrical power systems, etc. \cite{Dai89,Duan10,it17}. An interesting issue is therefore about properties of an NDS with the dynamics of its subsystems being described by descriptor forms.

In this section, complete observability and controllability are investigated for an NDS, assuming that the dynamics of its subsystems are described by a descriptor form with its system matrices being an LFT of some PFPPs, under the hypothesis that each subsystem and the whole NDS are well-posed. Results of the previous sections are extended.

More precisely, assume that in an NDS ${\bf \Sigma}^{[d]}$ composing of $N$ subsystems, the dynamics of its $i$-th subsystem, denoted by ${\bf \Sigma}_i^{[d]}$ with $i\in\{1,2,\cdots,N\}$, is described by the following descriptor form,
\begin{equation}
\hspace*{-0.25cm}\begin{array}{l}
\left[\!\! {\begin{array}{*{20}{c}}
E^{[0]}(i) \delta(x(t,i))\\
{{z}(t,i)}\\
{{y}(t,i)}
\end{array}}\!\! \right] \!=\! \left\{\! {\left[\!\! {\begin{array}{*{20}{c}}
{A_{\rm\bf xx}^{[0]}(i)} & {A_{\rm\bf xv}^{[0]}(i)} & {B_{\rm\bf x}^{[0]}(i)}\\
{A_{\rm\bf zx}^{[0]}(i)} & {A_{\rm\bf zv}^{[0]}(i)} & {B_{\rm\bf z}^{[0]}(i)}\\
{C_{\rm\bf x}^{[0]}(i)} & {C_{\rm\bf v}^{[0]}(i)} & {D^{[0]}(i)}
\end{array}}\!\! \right] \!\!+\!\! } \right.\\
\hspace*{0.2cm}\left.  \left[ \!\!\begin{array}{*{20}{l}}
{H_1{(i)}}\\
{H_2{(i)}}\\
{H_3{(i)}}
\end{array} \!\!\right]\!\!\! {P(i)}{{\left[I\!-\! {G{(i)}}{P{(i)}}\right]}^{ - 1}}\!\left[\!\! {\begin{array}{*{20}{c}}
{F_1{(i)}} \!\! & \!\! {F_2{(i)}} \!\! & \!\! {F_3{(i)}}
\end{array}}\!\! \right]\!\!\right\}\!\!\!\left[\!\!\! {\begin{array}{*{20}{c}}
{{x}(t,i)}\\
{{v}(t,i)}\\
{{u}(t,i)}
\end{array}}\!\!\! \right]
\end{array}
\label{eqn:46}
\end{equation}
in which $E^{[0]}(i)$ is a known square and real matrix that may not be invertible. In actual applications, this matrix is usually utilized to reflecting constraints on system states, etc. All the other matrices and vectors have the same meanings as those of Equation (\ref{eqn:1}). Moreover, relations among the internal subsystem inputs and outputs are still assumed to be described by Equation (\ref{eqn:2}).

To investigate controllability and observability of this NDS, some related concepts and results about a descriptor system are at first introduced, which are now well known \cite{Dai89,Duan10,it17}.

An LTI plant is called a descriptor system when its input-output relations are described by the following two equations,
\begin{equation}
E\delta(x(t))=Ax(t)+Bu(t), \hspace{0.5cm} y(t)=Cx(t)+Du(t)
\label{eqn:48}
\end{equation}
Here, $A$, $B$, $C$, $D$ and $E$ are some constant real matrices with compatible dimensions. If there exists a $\lambda\in{\cal C}$ such that ${\rm\bf det}(\lambda E - A)\neq 0$, then it is said to be regular. A descriptor system is said to be completely observable, if its initial states can be uniquely determined by its inputs and outputs over the whole time interval.

Regularity is particular and important to a descriptor system. When stimulated by a consistent input, a regular descriptor system has an unique output.

\begin{lemma}
Assume that the descriptor system described by Equation (\ref{eqn:48}) is regular. It is completely observable, if and only if the next two conditions are simultaneously satisfied,
\begin{itemize}
\item the matrix pencil $\left[\begin{array}{c} \lambda E - A \\ C \end{array}\right]$ is of FCR at every $\lambda\in{\cal C}$;
\item the matrix $\left[\begin{array}{c} E \\ C \end{array}\right]$ is of FCR.
\end{itemize}
\label{lemma:5}
\end{lemma}

Through completely the same augmentations as those of Section III for the internal input and output vectors of the subsystem ${\bf \Sigma}_i$ with $1\leq i\leq N$, the input-output relations of the NDS ${\bf \Sigma}^{[d]}$ can be equivalently rewritten as Equation (\ref{eqn:9}) and the following equation with $i\in\{1,2,\cdots,N\}$,
\begin{equation}
\left[ \begin{matrix}
E^{[0]}(i)\delta(x(t,i))\\
{{\bar{z}}(t,i)}\\
{{y}(t,i)}
\end{matrix} \right] = \left[ \begin{matrix}
{A_{\rm\bf xx}{(i)}} & {A_{\rm\bf xv}{(i)}} & {B_{\rm\bf x}{(i)}}\\
{A_{\rm\bf zx}{(i)}} & {A_{\rm\bf zv}{(i)}} & {B_{\rm\bf z}{(i)}}\\
{C_{\rm\bf x}{(i)}} & {C_{\rm\bf v}{(i)}} & {D{(i)}}
\end{matrix} \right]\left[ \begin{matrix}
{x(t,i)}\\
{\bar{v}(t,i)}\\
{u(t,i)}
\end{matrix} \right]
\label{eqn:47}
\end{equation}
Here, all the matrices have the same definitions as those of Equations (\ref{eqn:11}) and (\ref{eqn:9}).

Define a matrix $D_{\rm\bf u}$ as $D_{\rm\bf u}={\rm\bf diag}\left\{D^{[0]}(i)|_{i=1}^{N}\right\}$. Using this equivalent representation of the NDS ${\bf \Sigma}^{[d]}$, it can be straightforwardly shown through eliminating the internal input vector $\bar{v}(t)$ and the internal output vector $\bar{z}(t)$ that, its dynamics can also be described by the descriptor system of Equation (\ref{eqn:48}) with $E={\rm\bf diag}\left\{E^{[0]}(i)|_{i=1}^{N}\right\}$ and
\begin{equation}
\left[\!\!\begin{array}{cc} A & B \\  C & D
\end{array}\!\!\right] \!\!=\!\!\left[\!\!\begin{array}{cc}
A_{\rm\bf xx} & \hspace*{-0.2cm} B_{\rm\bf x} \\
C_{\rm\bf x}  & \hspace*{-0.2cm} D_{\rm\bf u} \end{array}\!\!\right]\!\!+\!\!\left[\!\!\begin{array}{c}
A_{\rm\bf xv} \\
C_{\rm\bf v}\end{array}\!\!\right]\!\!\bar{\Phi}\!
\left[\;I\!-\!A_{\rm\bf zv}\bar{\Phi}\;\right]^{\!-1}\!\left[
A_{\rm\bf zx}\;\; B_{\rm\bf z}\right]
\label{eqn:49}
\end{equation}
in which the matrices $A_{\rm\bf \# *}$, $B_{\rm\bf *}$ and $C_{\rm\bf *}$ with
${\rm\bf *,\#}={\rm\bf x}$, ${\rm\bf v}$, or ${\rm\bf z}$, have the same definitions as those of Lemma \ref{lemma:1}.

From the definition of regularity and the above equation, a condition can be established for the NDS ${\bf \Sigma}^{[d]}$ being regular.

\begin{theorem}
Denote a set consisting of $M_{\rm\bf x}+1$ arbitrary but different complex numbers by $\bar{\rm\bf\Lambda}$. Then the  NDS ${\rm\bf\Sigma}^{[d]}$ is regular if and only if there is a $\lambda_{0}\in{\bar{\rm\bf\Lambda}}$ such that the following matrix is of FCR,
\begin{equation}
\left[\begin{array}{cc}
\lambda_{0} E -A_{\rm\bf xx} & -A_{\rm\bf xv} \\
-\bar{\Phi} A_{\rm\bf zx} & I_{M_{\rm\bf z}}-\bar{\Phi} A_{\rm\bf zv} \end{array}\right]
\label{eqn:52}
\end{equation}
\label{theorem:6}
\end{theorem}

\hspace*{-0.4cm}{\it\bf Proof:}  For brevity, denote the matrix pencil of Equation (\ref{eqn:52}) by $\Pi(\lambda)$. When each subsystem of the NDS ${\rm\bf\Sigma}^{[d]}$ and the whole system are well-posed, it can be directly proven through matrix manipulations that the matrix $I_{M_{\rm\bf v}}-\bar{\Phi} A_{\rm\bf zv}$ is of FR \cite{zhou15,zyl18}. Hence
\begin{eqnarray}
{\rm\bf det}\left\{\Pi(\lambda)\right\}\! \!\!\!
&=&\!\!\!\! {\rm\bf det}\left(I_{M_{\rm\bf z}}-\bar{\Phi} A_{\rm\bf zv} \right)\times \nonumber \\
& & \hspace*{-0.2cm}
{\rm\bf det}\!\left(\!\! \lambda E \!-\!\left[ \! A_{\rm\bf xx} \!+\! A_{\rm\bf xv}\!\left[\! I_{M_{\rm\bf z}} \!-\! \bar{\Phi} A_{\rm\bf zv}\!\right]^{-1}\!\!\bar{\Phi} A_{\rm\bf zx}\!\right] \right) \nonumber\\
&=&\!\!\!\! {\rm\bf det}\left(I_{M_{\rm\bf z}}-\bar{\Phi} A_{\rm\bf zv}   \right)\times
{\rm\bf det}\left( \lambda E -A\right)
\label{eqn:53}
\end{eqnarray}

That is, at every $\lambda$, the nonsingularities of the matrix pencils $\Pi(\lambda)$ and $\lambda E -A$ are equivalent to each other.

Assume now that there is a $\lambda_{0}\in {\bar{\rm\bf\Lambda}}$ such that the matrix $\Pi(\lambda_{0})$ is of FCR. Then the above arguments mean that at least at this specific $\lambda_{0}$, the inverse of the matrix pencil $\lambda E-A$ exists. The regularity of the NDS ${\rm\bf\Sigma}^{[d]}$ follows directly from definitions.

On the contrary, assume that the matrix pencil $\Pi(\lambda)$ is not invertible for every $\lambda_{0}\in{\bar{\rm\bf\Lambda}}$. Then the nonsingularity equivalence between the matrices $\Pi(\lambda_{0})$ and  $\lambda_{0} E -A$ means that ${\rm\bf det}\left( \lambda_{0} E -A\right)\equiv 0$ whenever $\lambda_{0}\in{\bar{\rm\bf\Lambda}}$. Note that ${\rm\bf det}\left( \lambda E -A\right)$ is a polynomial having a degree not greater than $M_{\rm\bf x}$. Hence, if ${\rm\bf det}\left( \lambda E -A\right)\not\equiv 0$, then it has at most $M_{\rm\bf x}$ roots. Recall that the set ${\bar{\rm\bf\Lambda}}$ is formed by $M_{\rm\bf x}+1$ distinguished complex numbers. It can then be claimed that if ${\rm\bf det}\left(\Pi(\lambda_{0})\right)=0$ for every $\lambda_{0}\in{\bar{\rm\bf\Lambda}}$, then ${\rm\bf det}\left( \lambda E -A\right)\equiv 0$. That is, the NDS ${\rm\bf\Sigma}^{[d]}$ is not regular. This completes the proof.    \hspace{\fill}$\Diamond$

For a particular $\lambda_{0}$, the nonsingularity of the matrix $\Pi(\lambda_{0})$ can be verified through a matrix completely in the same form of Equation (\ref{eqn:32}). As a matter of fact, a direct application of Lemma \ref{lemma:3} with
\begin{displaymath}
\hspace*{-0.5cm} M_{1}\!=\!\left[\!\lambda_{0} E \!-\! A_{\rm\bf xx}\;\;-\! A_{\rm\bf xv} \!\right], \hspace{0.2cm}
M_{2}\!=\!\left[\!-\bar{\Phi} A_{\rm\bf zx} \;\; I_{M_{\rm\bf z}} \!-\! \bar{\Phi} A_{\rm\bf zv}\!\right]
\end{displaymath}
leads to this conclusion. On the other hand, note that the matrix $E$ and the matrix $A_{\rm\bf xx}$, as well as the matrix $A_{\rm\bf xv}$, have a consistent block diagonal structure. This means that in getting $\left[\lambda_{0} E \!\!-\!\! A_{\rm\bf xx}\;\; -\!\!A_{\rm\bf xv}\right]^{\perp}$, the associated computations can be done on each individual subsystem independently.

On the basis of Lemma \ref{lemma:5} and Equation (\ref{eqn:49}), the following results are gotten for NDS complete observability, which are very similar to those of Lemma \ref{lemma:1}. The latter is established in \cite{zhou15,zyl18} for an NDS with its subsystems being described by a state-space model.

\begin{theorem}
Assume that the NDS is regular. Define respectively a matrix $M^{[d]}_{\infty}$ and a matrix pencil $M^{[d]}(\lambda)$ as
\begin{eqnarray}
& & M^{[d]}(\lambda)=\left[\begin{array}{cc}
\lambda E -A_{\rm\bf xx} & -A_{\rm\bf xv} \\
-C_{\rm\bf x} & -C_{\rm\bf v} \\
-\bar{\Phi} A_{\rm\bf zx} & I_{M_{\rm\bf z}}-\bar{\Phi} A_{\rm\bf zv} \end{array}\right]
\label{eqn:50}   \\
& &
M^{[d]}_{\infty}= \left[\begin{array}{cc}
E  & 0 \\
-C_{\rm\bf x} & -C_{\rm\bf v} \\
-\bar{\Phi} A_{\rm\bf zx} & I_{M_{\rm\bf v}}-\bar{\Phi} A_{\rm\bf zv}
\end{array} \right]
\label{eqn:51}
\end{eqnarray}
Then the NDS ${\rm\bf\Sigma}^{[d]}$ is completely observable, if and only if the next two conditions are simultaneously satisfied,
\begin{itemize}
\item the matrix $M^{[d]}_{\infty}$ is of FCR;
\item at each $\lambda\in{\cal C}$, the matrix pencil $M^{[d]}(\lambda)$ is of FCR.
\end{itemize}
\label{theorem:5}
\end{theorem}

\hspace*{-0.4cm}{\it\bf Proof:}
For brevity, define a matrix pencil $\bar{M}^{[d]}(\lambda)$ and a matrix $\bar{M}^{[d]}_{\infty}$ respectively as
\begin{eqnarray*}
& &\hspace*{-0.5cm} \bar{M}^{[d]}(\lambda)\!=\!\left[\!\!\begin{array}{c} \lambda E -\left[A_{\rm\bf xx}+A_{\rm\bf xv}
(I_{M_{\rm\bf v}}-\bar{\Phi} A_{\rm\bf zv})^{-1}\bar{\Phi} A_{\rm\bf zx} \right] \\
C_{\rm\bf x}+C_{\rm\bf v}
(I_{M_{\rm\bf v}}-\bar{\Phi} A_{\rm\bf zv})^{-1}\bar{\Phi} A_{\rm\bf zx}\end{array} \!\!\right]   \\
& & \hspace*{-0.5cm}\bar{M}^{[d]}_{\infty}\!=\!\left[\begin{array}{c}  E  \\
C_{\rm\bf x}+C_{\rm\bf v}
(I_{M_{\rm\bf v}}-\bar{\Phi} A_{\rm\bf zv})^{-1}\bar{\Phi} A_{\rm\bf zx}\end{array}\right]
\end{eqnarray*}
Then according to Lemma \ref{lemma:5} and Equation (\ref{eqn:49}), the NDS ${\rm\bf\Sigma}^{[d]}$ is completely observable, if and only if $\bar{M}^{[d]}_{\infty}$ is of FCR and $\bar{M}^{[d]}(\lambda)$ is of FCR at an arbitrary $\lambda\in{\cal C}$.

When the NDS ${\bf \Sigma}^{[d]}$ is well-posed, we have that the matrix $I_{M_{\rm\bf v}}-\bar{\Phi} A_{\rm\bf zv}$ is invertible. Therefore
\begin{equation}
\left[-\bar{\Phi} A_{\rm\bf zx}\;\; I_{M_{\rm\bf v}} \!\!-\!\!\bar{\Phi} A_{\rm\bf zv}\right]^{\perp}={\rm\bf col}\left\{I,\;\;\bar{\Phi} A_{\rm\bf zx}(I_{M_{\rm\bf v}} \!\!-\!\!\bar{\Phi} A_{\rm\bf zv})^{-1}\right\}
\end{equation}
The proof can now be completed through an application of Lemma \ref{lemma:3}.    \hspace{\fill}$\Diamond$

Note that in the above theorem, except the replacement of the identity matrix $I_{M_{\rm\bf x}}$ in the matrix pencil $M(\lambda)$ of Lemma \ref{lemma:1} by the matrix $E$ in the matrix pencil $M^{[d]}(\lambda)$, these two matrix pencils are completely the same. On the other hand, note that the results of Lemma \ref{lemma:2} are valid for every matrix pencil with the form $\lambda G + H$. It is obvious that through similar arguments as those of Sections III and IV, similar conclusions can be obtained for the verification of the second condition of Theorem \ref{theorem:5}, and therefore the complete observability of an NDS whose subsystems being modeled by a descriptor form, as well as for its sensor placements.

Concerning verification of the condition associated with the matrix $M^{[d]}_{\infty}$, let $M_{1}$ and $M_{2}$ represent respectively the matrix
$\left[\begin{array}{cc}
E  & 0 \\
-C_{\rm\bf x} & -C_{\rm\bf v}
\end{array} \right]$
and the matrix $\left[-\bar{\Phi} A_{\rm\bf zx} \;\; I_{M_{\rm\bf v}} \!-\! \bar{\Phi} A_{\rm\bf zv}\right]$. Then a direct application of Lemma \ref{lemma:3} leads to an equivalent condition in the form of Equation (\ref{eqn:32}). Once again, except the augmented SCM $\bar{\Phi}$, the other two matrices in this condition can be obtained from independent computations on each individual subsystem.

In addition, based on the duality between complete controllability and complete observability of descriptor systems, similar results can be obtained for the complete controllability of the NDS ${\rm\bf\Sigma}^{[d]}$, as well as for its actuator placements.

\section{An Artificial Example}

To illustrate applicability of the obtained results in system analyses and syntheses, an artificial NDS is constructed and analyzed in this section, which has $N$ subsystems and each of them are constituted from two operational amplifiers, two capacitors and several resistors. Figure 1 gives a schematic illustration of its $i$-th subsystem with $1\leq i\leq N$.

The following assumptions are adopted for this system in which $i\in\{1,2,\cdots,N\}$.
\begin{itemize}
\item For each subsystem, the voltage of its right capacitor is measured.
\item The $i$-th subsystem is directly affected by $m(i)$ subsystems with their indices being $\rho_{1}(i)$, $\rho_{2}(i)$, $\cdots$, $\rho_{m(i)}(i)$.
\item The internal output of the $i$-th subsystem, that is, $z(t,i)$, directly affects $n(i)$ subsystems with their indices being $\xi_{1}(i)$, $\xi_{2}(i)$, $\cdots$, $\xi_{n(i)}(i)$.
\item The $i$-th subsystem directly affects the $\xi_{j}(i)$-th subsystem as its $\eta_{j}(i)$-th internal input, $1\leq j\leq n(i)$.
\end{itemize}
Define $\bar{R}_{i}$ for each $i=1,2,\cdots,N$ as
\begin{displaymath}
\bar{R}_{i}=\left(\sum_{j=1}^{n(i)}\frac{1}{R_{\xi_{j}(i),\;\eta_{j}(i)}}\right)^{-1}
\end{displaymath}
Moreover, denote $R_{i}C_{i}$ and $R_{i}/\bar{R}_{i}$ respectively by $T_{i}$ and $k_{i}$. In addition, let $x_{1}(t,i)$ and $x_{2}(t,i)$ represent respectively the voltages of the left and right capacitors in the $i$-th subsystem, and define its state vector  $x(t,i)$ as
$x(t,i)={\rm\bf col}\left\{x_{1}(t,i),\;x_{2}(t,i)\right\}$. Using these symbols, the following model can be straightforwardly established from circuit principles for the dynamics of this subsystem,
\begin{eqnarray*}
\dot{x}(t,i)\!\! &=&\!\! \frac{1}{(5+3k_{i})T_{i}}\left[\begin{array}{cc} -3-2k_{i} & 1 \\ 1 & -2-3k_{i} \end{array}\right]x(t,i)+  \nonumber\\
& & \frac{1}{(5+3k_{i})T_{i}}\left[\begin{array}{c} 2+k_{i} \\ 1 \end{array}\right]\left(v(t,i)-\frac{R_{i}^{*}}{R_{i,0}}u(t,i)\right)   \\
z(t,i)\!\! &=&\!\! \frac{1}{5+3k_{i}}\!\left\{\![1\;\; 3]x(t,i)+\left(\!\!v(t,i)-\frac{R_{i}^{*}}{R_{i,0}}u(t,i)\!\!\right)\!\!\right\}  \\
y(t,i)\!\! &=&\!\! [0\;\; 1]x(t,i)
\end{eqnarray*}
In addition, subsystem connections are given by the following equation
\begin{equation}
v(t,i)=\sum_{j=1}^{m(i)}\frac{R_{i}^{*}}{R_{i,j}}z(t,\rho_{j}(i))
\end{equation}
This implies that for each $1\leq j\leq m(i)$ and each $1\leq i\leq N$, the $i$-th row $\rho_{j}(i)$-column element of the SCM $\Phi$ is equal to $R_{i}^{*}/R_{i,j}$, while all the other elements are equal to zero.

Clearly, each element in the above system matrices, as well as the SCM, is a rational function of the physical parameters, which can be further expressed by an LFT. As this expression does not affect conclusions of this example, the details are not included. In addition, this NDS is well-posed if and only if
\begin{displaymath}
{\rm\bf det}\left(I-{\rm\bf diag}\left\{\left.\frac{1}{5+3k_{i}}\right|_{i=1}^{N}\right\}\Phi\right)\neq 0
\end{displaymath}

From this model, it can be directly proved that for each $i=1,2,\cdots,N$,
\begin{displaymath}
N_{\rm\bf cx}(i)=\left[\begin{array}{cc} 1 & 0 \\ 0 & 0 \end{array}\right], \hspace{0.5cm}
N_{\rm\bf cv}(i)=\left[0\;\; 1 \right]
\end{displaymath}
Using these matrices, direct algebraic manipulations show that
\begin{eqnarray*}
& & \left[\begin{array}{c}
\lambda N_{\rm\bf cx} -A_{\rm\bf xx}N_{\rm\bf cx}-A_{\rm\bf xv}N_{\rm\bf cv} \\
   N_{\rm\bf cv}-{\Phi} \left(A_{\rm\bf zx}N_{\rm\bf cx}+A_{\rm\bf zv}N_{\rm\bf cv}\right) \end{array}\right] \\
&=&
\left[\!\!\begin{array}{c}
{{\rm\bf diag}\!\left\{\!\left. {\left[\begin{array}{cc} \lambda+\frac{3+2k_{i}}{(5+3k_{i})T_{i}} &  -\frac{2+k_{i}}{(5+3k_{i})T_{i}} \\ -\frac{1}{(5+3k_{i})T_{i}} & -\frac{1}{(5+3k_{i})T_{i}}\end{array}\right]} \right|_{i=1}^{N}\!\right\}}   \\
{{\rm\bf diag}\!\left\{\!\left.\left[0\;\; 1\right]\right|_{i=1}^{N}\right\}}-\Phi
{{\rm\bf diag}\!\left\{\!\left. {\frac{1}{(5+3k_{i})T_{i}}\left[1 \;\; 1\right]} \right|_{i=1}^{N}\!\right\}}
\end{array}\!\!\!\right]
\end{eqnarray*}

On the other hand, for an arbitrary $i\in\{1,2,\cdots, N\}$,
\begin{displaymath}
{\left[\!\!\begin{array}{cc} \lambda+\frac{3+2k_{i}}{(5+3k_{i})T_{i}} &  -\frac{2+k_{i}}{(5+3k_{i})T_{i}} \\ -\frac{1}{(5+3k_{i})T_{i}} & -\frac{1}{(5+3k_{i})T_{i}}\end{array}\!\!\right]}
= U(i)
{\left[\!\!\begin{array}{cc} \lambda+T_{i}^{-1} & 0 \\ 0 & 1 \end{array}\!\!\!\right]}V(i)
\end{displaymath}
in which
\begin{displaymath}
U(i)={\left[\begin{array}{cc}1 & -\frac{2+k_{i}}{(5+3k_{i})T_{i}} \\ 0 & -\frac{1}{(5+3k_{i})T_{i}} \end{array}\!\!\!\right]}, \hspace{0.5cm}
V(i)={\left[\begin{array}{cc} 1 & 0 \\ 1 & 1 \end{array}\!\!\!\right]}
\end{displaymath}
Note that from their definitions, both $T_{i}$ and $k_{i}$ are positive numbers, which means that the matrices $U(i)$ and $V(i)$ are always invertible. Moreover,
\begin{eqnarray*}
& & N_{\rm\bf cv}(i)V_{i}^{-1}(1:1)=-1 \\
& & \left(A_{\rm\bf zx}(i)N_{\rm\bf cx}(i)+A_{\rm\bf zv}(i)N_{\rm\bf cv}(i)\right)V_{i}^{-1}(1:1)=0
\end{eqnarray*}

\renewcommand{\thefigure}{\arabic{figure}}
\setcounter{figure}{0}
\vspace{-0.0cm}
\begin{figure}[!ht]
\begin{center}
\includegraphics[width=3.0in]{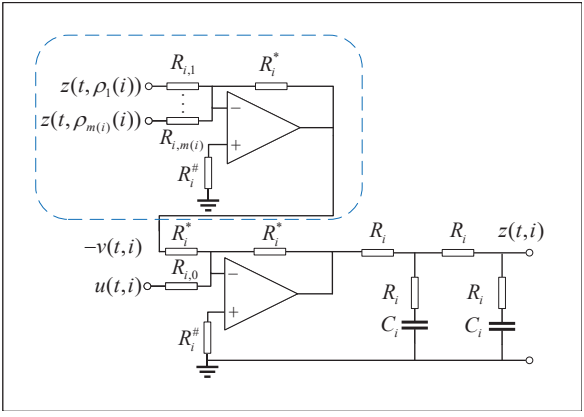}
\vspace{-0.2cm}\hspace*{5cm} \caption{The $i$-th Subsystem of the NDS}
\end{center}
\label{figure:1}
\end{figure}

Let $\rm\bf\Lambda$ denote the set constituted from $-1/T_{i}|_{i=1}^{N}$ with different values. For each $\lambda_{0}\in {\rm\bf\Lambda}$, let ${\cal M}(\lambda_{0})$ denote the set of all the indices of subsystems with $T_{i}=-1/\lambda_{0}$. Then
\begin{displaymath}
\bar{\Xi}^{\perp}(\lambda_{0})={\rm\bf col}\left\{\begin{array}{c} 0 \\ {\rm\bf diag}\left\{\left.
{\left[\!\!\begin{array}{c} 1 \\ 0 \end{array}\!\!\right]}\right|_{j\in {\cal M}(\lambda_{0})} \right\} \end{array}\right\}
\end{displaymath}
in which the zero vectors usually have different dimensions. Substitute these results into the definitions of the matrices $X(\lambda_{0})$ and $X(\lambda_{0})$ of Theorem \ref{theorem:3},we immediately have that
\begin{displaymath}
X(\lambda_{0})=-\bar{\Xi}^{\perp}(\lambda_{0}), \hspace{0.5cm} Y(\lambda_{0})=0
\end{displaymath}
Hence $X(\lambda_{0})-\Phi Y(\lambda_{0})\equiv -\bar{\Xi}^{\perp}(\lambda_{0})$, and is therefore always of FCR, no matter what value $\lambda_{0}$ takes from the set $\rm\bf\Lambda$ and what value the SCM $\Phi$ takes.

It can therefore be declared that when this artificial NDS is well-posed, it is always observable, no matter how its subsystems are connected.

When the NDS is well-posed and the voltage of the left capacitor is measured for each subsystem, similar arguments show that the system is observable, if and only if the matrix $I+\Phi$ is invertible.

Using the results of the previous sections, this system can be proved to be also controllable for arbitrary subsystem connections, provided that it is well-posed.

It is worthwhile to emphasize that in large scale NDS analysis and synthesis, computational complexity is a very important issue. Note that the results obtained in this paper, such as Theorem \ref{theorem:3}, have  completely the same form as Theorem 2 of \cite{zhou15}. It can be claimed that they share the same computational properties. That is, their computation costs increase linearly with the subsystem number $N$, while those using a lumped model increase at least in the order of $N^{3}$. Moreover, computations with these results are numerically more stable. Analysis details, as well as some numerical simulation comparisons, can be found in \cite{zhou15}. 

\section{Concluding Remarks}

This paper studies observability/controllability of networked dynamic systems, in which the system matrices of each subsystem are expressed through a linear fractional transformation of its (pseudo) first principle parameters. Some explicit connections have been obtained between an NDS and a descriptor system in their observability/controllability. By means of the Kronecker canonical form of matrix pencils, a rank based necessary and sufficient condition is gotten with the associated matrix depending affinely on subsystem parameters/connections. While the corresponding diagonal blocks are computed in a significantly different way, this matrix has a form that completely agrees with those of \cite{zhou15,zyl18}. In addition, in its derivations, the full normal rank condition asked there is no longer required. On the other hand, this matrix keeps the attractive property that in getting the associated matrices, all the numerical calculations are performed on each individual subsystem independently. This makes the condition verification scalable for a large scale NDS. Moreover, except well-posedness of the whole system and its subsystems, there are not any other restrictions on either a subsystem  or the subsystem connections.

On the basis of this condition, characteristics of a subsystem are clarified with which an observable/controllable NDS can be more easily constructed. It has been made clear that
satisfaction of the full normal column/row rank condition adopted in \cite{zhou15,zyl18} may greatly reduce difficulties in constructing an observable/controllable NDS. In addition, subsystems with an input matrix of full column rank are helpful in constructing an observable NDS that receives signals from other subsystems, while subsystems with an output matrix of full row rank are helpful in constructing a controllable NDS that sends signals to other subsystems. Extensions to an NDS whose subsystems are  modeled by a descriptor form have also been attacked. It is observed that similar results can also be established.

Further efforts include finding engineering significant explanations for the obtained results, as well as extending the obtained results to structural controllability/observability of a networked dynamic system.

\renewcommand{\theequation}{a.\arabic{equation}}
\setcounter{equation}{0}

\begin{appendices}

\section{Proof of Lemma \ref{lemma:0}}

In this proof, the null space is derived for the associated 5 kinds of matrix pencils individually.

\subsection{The Matrix Pencil $H_{m}(\lambda)$}

Note that the matrix pencil $H_{m}(\lambda)$ is strictly regular. According to the definition of a strictly regular matrix pencil, there are two real and nonsingular $m\times m$ dimensional matrices $X$ and $Y$, such that $H_{m}(\lambda)=\lambda X+Y$. Hence,
\begin{eqnarray*}
& & H_{m}(\lambda)\alpha =X\left[\lambda I+X^{-1}Y\right]\alpha \\
& &
{\rm\bf det}\left[H_{m}(\lambda)\right]={\rm\bf det}\left[X\right]\times {\rm\bf det}\left[\lambda I+X^{-1}Y\right]
\end{eqnarray*}
in which $\alpha$ is an arbitrary vector with a consistent dimension.
It can therefore be declared that ${\rm\bf det}\left[H_{m}(\lambda)\right]=0$ if and only if ${\rm\bf det}\left[\lambda I+X^{-1}Y\right]=0$. On the other hand, note that
\begin{displaymath}
{\rm\bf det}\left[X^{-1}Y\right]=\frac{{\rm\bf det}\left[Y\right]}{{\rm\bf det}\left[X\right]}
\end{displaymath}
which is different from zero by the nonsingularity of the associated two matrices. We therefore have that the matrix $X^{-1}Y$ is always nonsingular. Moreover, $H_{m}(\lambda)\alpha=0$ if and only if $\left[\lambda I+X^{-1}Y\right]\alpha=0$. That is, the matrix pencil $H_{m}(\lambda)$ is rank deficient only at eigenvalues of the matrix $X^{-1}Y$ which are isolated and different from zero.

\subsection{The Matrix Pencils $N_{m}(\lambda)$ and $J_{m}(\lambda)$ }

From the definitions of these two matrix pencils, direct algebraic manipulations show that
\begin{displaymath}
{\rm\bf det}\left[N_{m}(\lambda)\right]\equiv 1
\end{displaymath}
On the other hand, if there is a vector $\alpha=[\alpha_{1}\; \alpha_{2}\; \cdots\; \alpha_{m}]^{T}$, such that $J_{m}(\lambda)\alpha=0$ is satisfied at a particular value of the complex variable $\lambda$, say, $\lambda_{0}$. Then
\begin{eqnarray*}
& & \lambda_{0}\alpha_{1}=0 \\
& & \alpha_{i}+\lambda_{0}\alpha_{i+1}=0,\hspace{0.25cm} i=1,2,\cdots,m-1 \\
& & \alpha_{m}=0
\end{eqnarray*}
which lead to $\alpha_{1}=\alpha_{2}=\cdots=\alpha_{m}=0$. That is, $\alpha=0$. Hence, the matrix $J_{m}(\lambda)$ is of FCR, no matter what value is taken by the complex variable $\lambda$.

\subsection{The Matrix Pencil $K_{m}(\lambda)$}

Assume that there is a vector $\alpha=[\alpha_{1}\; \alpha_{2}\; \cdots\; \alpha_{m}]^{T}$, such that $K_{m}(\lambda)\alpha=0$ is satisfied at a particular value of the complex variable $\lambda$. Denote it by $\lambda_{0}$. Then
\begin{eqnarray}
& & \lambda_{0}\alpha_{i}+\alpha_{i+1}=0,\hspace{0.25cm} i=1,2,\cdots,m-1
\label{eqn:17}\\
& & \lambda_{0}\alpha_{m}=0
\end{eqnarray}

If $\lambda_{0}\neq 0$, then the last equation means that $\alpha_{m}=0$,
which further leads to that $\alpha_{m-1}=\alpha_{{m}-2}=\cdots=\alpha_{1}=0$. Hence, the matrix pencil $K_{m}(\lambda)$ is always of FCR whenever $\lambda\neq 0$.
On the other hand, if $\lambda_{0}= 0$, then Equation (\ref{eqn:17}) implies that $\alpha_{m}=\alpha_{{m}-1}=\cdots=\alpha_{2}=0$, while $\alpha_{1}$ can be an arbitrary complex number, which is the result needed to prove.

\subsection{The Matrix Pencil $L_{m}(\lambda)$}

Assume that there is a vector $\alpha=[\alpha_{1}\; \alpha_{2}\; \cdots\; \alpha_{m}]^{T}$, such that $L_{m}(\lambda)\alpha=0$ is satisfied with  $\lambda=\lambda_{0}$, a particular value of the complex variable $\lambda$. Then
\begin{equation}
\lambda_{0}\alpha_{i}+\alpha_{i+1}=0,\hspace{0.25cm} i=1,2,\cdots,m
\label{eqn:20}
\end{equation}

If $\lambda_{0}\neq 0$, then Equation (\ref{eqn:20}) means that $\alpha_{i+1}=-\lambda_{0}\alpha_{i}$, $i=1,2,\cdots,m$. On the other hand,
if $\lambda_{0}= 0$, this equation implies that $\alpha_{m}=\alpha_{{m}-1}=\cdots=\alpha_{2}=0$, while $\alpha_{1}$ can take any complex value. These prove the results.     \hspace{\fill}$\Diamond$

\section{Proof of Theorem 1}

For brevity, define sets ${\cal K}$ and ${\cal N}$ respectively as
\begin{displaymath}
{\cal K}=\left\{\;k(1),\; k(2),\;\cdots,\;k(M)\;\right\}, \hspace{0.5cm}
{\cal N}=\left\{\;1,\; 2,\;\cdots,\;N\;\right\}
\end{displaymath}
Let $\alpha_{{\rm\bf x}}(i)$ be an arbitrary $m_{{\rm\bf x}i}$ dimensional real vector, while $\alpha_{{\rm\bf v}}(i)$ be an arbitrary $m_{{\rm\bf v}i}$ dimensional real vector, $i=1,2,\cdots,N$. Denote ${\rm\bf col}\!\left\{\alpha_{\rm\bf x}(i)|_{i=1}^{N},\;\alpha_{\rm\bf v}(i)|_{i=1}^{N}\right\}$ by $\alpha$. From the block diagonal structure of the matrices $C_{\rm\bf x}$ and $C_{\rm\bf v}$, it is immediate that
\begin{equation}
\left[ C_{\rm\bf x}\;\; C_{\rm\bf v} \right]\alpha
=\left[\begin{array}{c}
C_{\rm\bf x}(1)\alpha_{{\rm\bf x}}(1)+ C_{\rm\bf v}(1)\alpha_{{\rm\bf v}}(1) \\
C_{\rm\bf x}(2)\alpha_{{\rm\bf x}}(2)+ C_{\rm\bf v}(2)\alpha_{{\rm\bf v}}(2) \\
\vdots \\
C_{\rm\bf x}(N)\alpha_{{\rm\bf x}}(N)+ C_{\rm\bf v}(N)\alpha_{{\rm\bf v}}(N) \end{array}\right]
\end{equation}
It can therefore be declared that $\left[C_{\rm\bf x}\;\; C_{\rm\bf v}\right]\alpha=0$ if and only if
\begin{equation}
C_{\rm\bf x}(i)\alpha_{{\rm\bf x}}(i)+ C_{\rm\bf v}(i)\alpha_{{\rm\bf v}}(i)=0, \hspace{0.5cm} i=1,2,\cdots,N
\label{eqn:a10}
\end{equation}

From the adopted assumption, we have that when $i\in {\cal K}$, the matrix $[C_{\rm\bf x}(i)\;\;C_{\rm\bf v}(i)]$ is not of FCR, otherwise it is of FCR. Hence, Equation (\ref{eqn:a10}) implies that for each $j\in \{1,2,\cdots,M\}$, there is a vector $\xi(j)$, such that
\begin{equation}
\left[\begin{array}{c}
\alpha_{{\rm\bf x}}(k(j)) \\
\alpha_{{\rm\bf v}}(k(j))  \end{array}\right]=
\left[\begin{array}{c} N_{\rm\bf cx}(k(j))  \\ N_{\rm\bf cv}(k(j)) \end{array}\right]\xi(j)
\label{eqn:22}
\end{equation}
Moreover, $\alpha_{{\rm\bf x}}(i)=0$ and $\alpha_{{\rm\bf v}}(i)=0$ for each $i\in{\cal N}\backslash {\cal K}$. Denote the vector
${\rm\bf col}\left\{\xi(j)|_{j=1}^{M} \right\}$ by $\xi$. These results mean that
\begin{eqnarray}
\alpha &=&\! \!\left[\!\!\!\!\begin{array}{c} {\left[\!\!\begin{array}{c} 0  \\ {{\rm\bf col}\left\{\left. {\left[\begin{array}{c} N_{\rm\bf cx}(k(j))\xi(j) \\ 0 \end{array}\right]} \right|_{j=1}^{M}\right\}}\end{array}\!\!\right]}  \\
\\
{\left[\!\!\begin{array}{c} 0  \\ {{\rm\bf col}\left\{\left. {\left[\begin{array}{c} N_{\rm\bf cv}(k(j))\xi(j) \\ 0 \end{array}\right]} \right|_{j=1}^{M}\right\}}\end{array}\!\!\right]} \end{array}\!\!\!\!\right]\nonumber\\
&=&
\!\! \left[\begin{array}{c} N_{\rm\bf cx}  \\
N_{\rm\bf cv} \end{array}\right]\xi
\label{eqn:23}
\end{eqnarray}

On the other hand, a repetitive application of Lemma \ref{lemma:6} and Corollary \ref{corollary:2} shows that the matrix ${\rm\bf col}\left\{N_{\rm\bf cx},\; N_{\rm\bf cv}\right\}$ is of FCR. It can therefore be declared that
\begin{equation}
[C_{\rm\bf x} \;\;C_{\rm\bf v}]^{\perp}
={\rm\bf col}\left\{N_{\rm\bf cx},\;\; N_{\rm\bf cv}\right\}
\label{eqn:24}
\end{equation}

Note that rearranging rows of a matrix does not affect its rank. The proof can now be completed using Lemma \ref{lemma:3}.   \hspace{\fill}$\Diamond$

\section{Proof of Theorem 2}

Define a matrix pencil $\Xi(\lambda)$ as
\begin{displaymath}
\Xi(\lambda)=\left[\begin{array}{c} 0 \\ {\rm\bf diag}\!\left\{\left.{\rm\bf col}\!\left\{\Xi(\lambda,k(i)),\;0\right\}\right|_{i=1}^{M}\right\} \end{array}\right]
\end{displaymath}
From the compatible block diagonal structures of the matrices $N_{\rm\bf cx}$, $A_{\rm\bf xx}$ and $A_{\rm\bf xv}$, it is obvious that
\begin{eqnarray}
& & \lambda N_{\rm\bf cx}-\left[A_{\rm\bf xx}N_{\rm\bf cx}+A_{\rm\bf xv}N_{\rm\bf cv}\right] \nonumber\\
&=&
{\rm\bf diag}\!\left\{I,\; \left.{\rm\bf diag}\!\left\{U(k(i)),\;I\right\}\right|_{i=1}^{M}\right\}\times \nonumber\\
& & \hspace*{2cm} \Xi(\lambda) {\rm\bf diag}\!\left\{\left.V(k(i))\right|_{i=1}^{M}\right\}
\label{eqn:14-a}
\end{eqnarray}
In these expressions, the zero matrices and the identity matrices may have different dimensions. As their actual values are not very important in the following derivations, they are omitted for simplicity.

To simplify the associated expressions, denote the block diagonal matrices ${\rm\bf diag}\!\left\{I, \left.{\rm\bf diag}\!\left\{U(k(i)),\;I\right\}\right|_{i=1}^{M}\right\}$ and ${\rm\bf diag}\left\{V(k(i))|_{i=1}^{M}\right\}$ respectively by $U$ and $V$. Then both $U$ and $V$ are invertible. On the other hand, it is obvious from its definition that the matrix pencil $\Xi(\lambda)$ is block diagonal, and consists only of strictly regular matrix pencils, the matrix pencils in the forms of $K_{*}(\lambda)$, $N_{*}(\lambda)$, $L_{*}(\lambda)$ and $J_{*}(\lambda)$.

On the basis of Equations (\ref{eqn:3-a}) and (\ref{eqn:14-a}), we have that
\begin{eqnarray}
\Psi(\lambda)\!\!\!\!\!&=&\!\!\!\!\!\!\!
\left[\begin{array}{c}
U\Xi(\lambda)V \\
   N_{\rm\bf cv}-\bar{\Phi} \left(A_{\rm\bf zx}N_{\rm\bf cx}+A_{\rm\bf zv}N_{\rm\bf cv}\right) \end{array}\right]  \nonumber\\
&=&\!\!\!\!\!\!\!
\left[\!\!\begin{array}{cc}
U & 0  \\ 0 & I \end{array}\!\!\right] \tilde{\Psi}(\lambda) V
\label{eqn:30}
\end{eqnarray}
in which
\begin{displaymath}
\tilde{\Psi}(\lambda) =\left[\begin{array}{c}
 \Xi(\lambda) \\
   N_{\rm\bf cv}V^{-1}\!-\!\bar{\Phi} \left(A_{\rm\bf zx}N_{\rm\bf cx}V^{-1}+A_{\rm\bf zv}N_{\rm\bf cv}V^{-1}\right) \end{array}\right]
\end{displaymath}
As both the matrices $U$ and $V$ are invertible, it is obvious from Equation (\ref{eqn:30}) that the matrix pencil $\Psi(\lambda)$ is of FCR at every complex $\lambda$, if and only if the matrix pencil $\tilde{\Psi}(\lambda)$ holds this property.

Note that deleting zero rows of a matrix does not change its rank. Moreover, the matrix pencils $N_{*}(\lambda)$ and $J_{*}(\lambda)$ are always of FCR. On the basis of Lemma \ref{lemma:6}, this characteristic of $N_{*}(\lambda)$ and $J_{*}(\lambda)$ in the matrix pencil $\Xi(\lambda)$, as well as its block diagonal structure, it can be claimed that the matrix pencil $\tilde{\Psi}(\lambda)$ is always of FCR, if and only if the matrix pencil $\bar{\Psi}(\lambda)$ meets this requirement. This completes the proof.   \hspace{\fill}$\Diamond$

\section{Proof of Theorem 4}

For brevity, denote $\lambda N_{\rm\bf cx}(i)-\left[A_{\rm\bf xx}(i)N_{\rm\bf cx}(i)+\right.$  $\left.A_{\rm\bf xv}(i)N_{\rm\bf cv}(i)\right]$ by $\Omega(\lambda,i)$. From the assumption that $C_{\rm\bf v}(i)=0$, straightforward matrix operations show that
\begin{equation}
\left[C_{\rm\bf x}(i)\;\; C_{\rm\bf v}(i)\right]^{\perp}={\rm\bf diag}\left\{C_{\rm\bf x}^{\perp}(i),\; I \right\}
\label{eqn:38}
\end{equation}
From this relation, as well as the definitions of the matrices $N_{\rm\bf cx}(i)$ and $N_{\rm\bf cv}(i)$, we have that
\begin{equation}
N_{\rm\bf cx}(i)=\left[C_{\rm\bf x}^{\perp}(i) \;\; 0 \right],\hspace{0.5cm}
N_{\rm\bf cv}(i)=\left[0 \;\; I \right]
\label{eqn:39}
\end{equation}
Substitute these two equalities into the definition of the matrix pencil $\Omega(\lambda,i)$, we have that
\begin{equation}
\Omega(\lambda,i)=\left[(\lambda I-A_{\rm\bf xx}(i))C_{\rm\bf x}^{\perp}(i) \;\;
A_{\rm\bf xv}(i)\right]
\label{eqn:40}
\end{equation}

Assume that $A_{\rm\bf xv}(i)$ is not of FCR. Then there exists a nonzero vector $\alpha_{\rm\bf v}$, denote it by $\alpha_{{\rm\bf v}0}$ such that
$A_{\rm\bf xv}(i)\alpha_{{\rm\bf v}0}=0$. Construct a vector $\alpha_{0}$ as
$\alpha_{0}={\rm\bf col}\left\{0,\;\; \alpha_{{\rm\bf v}0}\right\}$ in which the zero vector has a suitable dimension. Obviously, this vector is not a zero vector. On the other hand, from Equation (\ref{eqn:40}), direct algebraic manipulations show that
$\Omega(\lambda,i)\alpha_{0}=0$ for an arbitrary complex $\lambda$. This means that the matrix pencil $\Omega(\lambda,i)$ is not of FNCR. Hence, to guarantee that this matrix pencil is of FNCR, it is necessary that the matrix $A_{\rm\bf xv}(i)$ is of FCR.

Assume now that the matrix pencil $\Omega(\lambda,i)$ is not of FNCR. Then for each  $\lambda_{0}\in{\cal C}$, the matrix $\Omega(\lambda_{0},i)$ is not of FCR. Let $\alpha$ be a nonzero complex vector satisfying $\Omega(\lambda_{0},i)\alpha=0$. Partition it as $\alpha={\rm\bf col}\left\{\alpha_{\rm\bf x},\;\; \alpha_{\rm\bf v}\right\}$ in a consistent way as the partition of the matrix pencil $\Omega(\lambda,i)$. Then
\begin{eqnarray}
\Omega(\lambda_{0},i)\alpha&=&(\lambda_{0} I-A_{\rm\bf xx}(i))C_{\rm\bf x}^{\perp}(i)\alpha_{\rm\bf x}+ A_{\rm\bf xv}(i)\alpha_{\rm\bf v} \nonumber\\
&=&0
\label{eqn:41}
\end{eqnarray}

When the matrix $A_{\rm\bf xv}(i)$ is of FCR, the matrix $A_{\rm\bf xv}^{T}(i)A_{\rm\bf xv}(i)$ is invertible. Moreover, Equation (\ref{eqn:41}) implies that
\begin{equation}
\alpha_{\rm\bf v}=-\left[A_{\rm\bf xv}^{T}(i)A_{\rm\bf xv}(i)\right]^{-1}A_{\rm\bf xv}^{T}(i)(\lambda_{0} I-A_{\rm\bf xx}(i))C_{\rm\bf x}^{\perp}(i)\alpha_{\rm\bf x}
\label{eqn:42}
\end{equation}
$\alpha_{\rm\bf x} \neq 0$ is hence guaranteed by $\alpha \neq 0$.

Recall that the matrix pencil $\Theta(\lambda,i)$ is defined as
$\Theta(\lambda,i)=\left\{I-A_{\rm\bf xv}(i)\left[A_{\rm\bf xv}^{T}(i)A_{\rm\bf xv}(i)\right]^{-1}A_{\rm\bf xv}^{T}(i)\right\}(\lambda I-A_{\rm\bf xx}(i))C_{\rm\bf x}^{\perp}(i)$. Substitute the above equality back into Equation (\ref{eqn:41}), we further have that
\begin{equation}
\Theta(\lambda_{0},i)\alpha_{\rm\bf x}=0
\label{eqn:43}
\end{equation}
Hence, $\alpha_{\rm\bf x} \neq 0$ and Equation (\ref{eqn:43}) imply that the matrix $\Theta(\lambda_{0},i)$ is not of FCR. Note that $\lambda_{0}$ is an arbitrary complex number. From the definition of FNCR, it can be declared that the matrix pencil $\Theta(\lambda,i)$ is not of FNCR.

On the contrary, assume that the matrix pencil $\Theta(\lambda,i)$ is not of FNCR. Then for an arbitrary $\lambda_{0}\in{\cal C}$, there is at least one vector $\alpha_{{\rm\bf x}0}$ that is nonzero and satisfies $\Theta(\lambda_{0},i)\alpha_{{\rm\bf x}0}=0$. Construct a vector $\alpha$ as
\begin{displaymath}
\alpha =\left[\begin{array}{c} \alpha_{{\rm\bf x}0} \\
-\left[A_{\rm\bf xv}^{T}(i)A_{\rm\bf xv}(i)\right]^{-1}A_{\rm\bf xv}^{T}(i)(\lambda I-A_{\rm\bf xx}(i))C_{\rm\bf x}^{\perp}(i)\alpha_{{\rm\bf x}0}
\end{array}\right]
\end{displaymath}
Clearly $\alpha\neq 0$. On the other hand, direct matrix operations lead to
\begin{equation}
\Omega(\lambda_{0},i)\alpha=\Theta(\lambda_{0},i)\alpha_{{\rm\bf x}0}=0
\label{eqn:44}
\end{equation}
which means that the matrix $\Omega(\lambda_{0},i)$ is also not of FCR. As $\lambda_{0}$ is an arbitrary complex number, this implies that $\Omega(\lambda,i)$ is always of rank deficient, and hence is not of FNCR.

This completes the proof.   \hspace{\fill}$\Diamond$

\end{appendices}

\end{document}